\documentclass[aps,prl,twocolumn,longbibliography,superscriptaddress,notitlepage,floatfix]{revtex4-2}
\renewcommand{\figurename}{Fig.}
\usepackage{amsmath,amsfonts,amssymb,color,xcolor,epsfig,graphics,graphicx,latexsym,theorem,url,multirow,diagbox}
\usepackage[unicode=true,bookmarks=true,bookmarksnumbered=false,bookmarksopen=false,breaklinks=false,pdfborder={0 0 1},backref=false,colorlinks=true]{hyperref}
\usepackage[utf8]{inputenc}
\usepackage{booktabs}
\usepackage{tabularx}
\usepackage[table,xcdraw]{xcolor}
\usepackage[T1]{fontenc}
\usepackage{courier}
\usepackage{listings}
\usepackage{latexsym}
\usepackage{dcolumn}
\usepackage{comment}
\usepackage{times} 
\usepackage{enumitem}
\usepackage[normalem]{ulem}
\usepackage{algorithm}
\usepackage{algorithmicx}
\usepackage{algpseudocode}

\usepackage{braket}
\usepackage{bm}

\usepackage{lineno}

\usepackage{soul}
\usepackage{xcolor}

\hypersetup{
 linkcolor=magenta, urlcolor=blue, citecolor=blue, pdfstartview={FitH}, unicode=true}

\makeatletter
\newcommand{\beginExtendedData}{%
    \newcounter{ExFig}
    \makeatletter
    \@addtoreset{figure}{ExFig}
    \makeatother
    \stepcounter{ExFig}
    \renewcommand{\figurename}{Extended Data Fig.}
   }

\makeatother

\begin{document}
\title{Combinatorial optimization enhanced by shallow quantum circuits with 104 superconducting qubits}

\author{Xuhao Zhu}\thanks{These authors contributed equally}
\affiliation{School of Physics, ZJU-Hangzhou Global Scientific and Technological Innovation Center, and Zhejiang Key Laboratory of Micro-nano Quantum Chips and Quantum Control, Zhejiang University, Hangzhou 310000, China}

\author{Zuoheng Zou}\thanks{These authors contributed equally}\affiliation{Huawei Technologies Co., Ltd, Shenzhen 518000, China}

\author{Feitong Jin}\thanks{These authors contributed equally}\affiliation{School of Physics, ZJU-Hangzhou Global Scientific and Technological Innovation Center, and Zhejiang Key Laboratory of Micro-nano Quantum Chips and Quantum Control, Zhejiang University, Hangzhou 310000, China}

\author{Pavel Mosharev}\thanks{These authors contributed equally}\affiliation{Huawei Technologies Co., Ltd, Shenzhen 518000, China}

\author{Maolin Luo}\affiliation{Huawei Technologies Co., Ltd, Shenzhen 518000, China}

\author{Yaozu Wu}

\author{Jiachen Chen}

\author{Chuanyu Zhang}

\author{Yu Gao}

\author{Ning Wang}

\author{Yiren Zou}

\author{Aosai Zhang}

\author{Fanhao Shen}

\author{Zehang Bao}

\author{Zitian Zhu}

\author{Jiarun Zhong}

\author{Zhengyi Cui}

\author{Yihang Han}

\author{Yiyang He}

\author{Han Wang}

\author{Jia-Nan Yang}

\author{Yanzhe Wang}

\author{Jiayuan Shen}

\author{Gongyu Liu}

\author{Zixuan Song}

\author{Jinfeng Deng}

\author{Hang Dong}

\author{Pengfei Zhang}

\affiliation{School of Physics, ZJU-Hangzhou Global Scientific and Technological Innovation Center, and Zhejiang Key Laboratory of Micro-nano Quantum Chips and Quantum Control, Zhejiang University, Hangzhou 310000, China}

\author{Chao Song}
\affiliation{School of Physics, ZJU-Hangzhou Global Scientific and Technological Innovation Center, and Zhejiang Key Laboratory of Micro-nano Quantum Chips and Quantum Control, Zhejiang University, Hangzhou 310000, China}

\affiliation{Hefei National Laboratory, Hefei 230088, China}

\author{Zhen Wang}

\affiliation{School of Physics, ZJU-Hangzhou Global Scientific and Technological Innovation Center, and Zhejiang Key Laboratory of Micro-nano Quantum Chips and Quantum Control, Zhejiang University, Hangzhou 310000, China}

\affiliation{Hefei National Laboratory, Hefei 230088, China}

\author{Hekang Li}

\affiliation{School of Physics, ZJU-Hangzhou Global Scientific and Technological Innovation Center, and Zhejiang Key Laboratory of Micro-nano Quantum Chips and Quantum Control, Zhejiang University, Hangzhou 310000, China}

\affiliation{Hefei National Laboratory, Hefei 230088, China}

\author{Qiujiang Guo}

\affiliation{School of Physics, ZJU-Hangzhou Global Scientific and Technological Innovation Center, and Zhejiang Key Laboratory of Micro-nano Quantum Chips and Quantum Control, Zhejiang University, Hangzhou 310000, China}

\affiliation{Hefei National Laboratory, Hefei 230088, China}

\author{Man-Hong Yung}
\email{yung.manhong@huawei.com}
\affiliation{Huawei Technologies Co., Ltd, Shenzhen 518000, China}
\author{H. Wang}
\email{hhwang@zju.edu.cn}
\affiliation{School of Physics, ZJU-Hangzhou Global Scientific and Technological Innovation Center, and Zhejiang Key Laboratory of Micro-nano Quantum Chips and Quantum Control, Zhejiang University, Hangzhou 310000, China}

\affiliation{Hefei National Laboratory, Hefei 230088, China}

\begin{abstract}
\textbf{
A pivotal task for quantum computing is to speed up solving problems that are both classically intractable and practically valuable. Among these, combinatorial optimization problems have attracted tremendous attention due to their broad applicability and natural fitness to Ising Hamiltonians.
Here we propose a quantum sampling strategy, based on which we design an algorithm for accelerating solving the ground states of Ising model, a class of NP-hard problems in combinatorial optimization. 
The algorithm employs a hybrid quantum-classical workflow, with a shallow-circuit quantum sampling subroutine dedicated to navigating the energy landscape. 
Using up to 104 superconducting qubits, we demonstrate that this algorithm outputs favorable solutions against even a highly-optimized classical simulated annealing (SA) algorithm. 
Furthermore, 
we illustrate the path toward quantum speedup based on the time-to-solution metric against SA running on a single-core CPU with just 100 qubits.
Our results indicate a promising alternative to classical heuristics for combinatorial optimization, a paradigm where quantum advantage might become possible on near-term superconducting quantum processors with thousands of qubits and without the assistance of error correction.
}
\end{abstract}

\maketitle

Combinatorial optimization is prevalent in science and engineering. 
Many such problems can be effectively mapped to solving the ground states of the Ising model, which generally belong to the complexity class of NP-hard~\cite{Barahona1982,Lucas2014}. 
The difficulty originates from the exponentially large solution space and the non-convexity of the corresponding energy landscape, 
which frequently traps optimizers in local minima.
Classical escape strategies typically involve flipping a subset of spins in the current solution~\cite{Kirkpatrick1983, Glover1989, Glover1990, Benlic2013}. 
However, these perturbations face a fundamental trade-off: if too weak, they fail to escape the local energy basin; if too strong, they disrupt the algorithm’s memory, akin to a random restart.
Despite extensive research, developing an efficient solver for the Ising model remains an ongoing challenge.

Quantum computing with its exploding computational space offers a natural platform for encoding and solving the Ising problems.
Capable of querying the energy function in quantum superposition, a quantum computer can propose new solutions based on the entire energy landscape, 
which facilitates efficient escape from energy basins through quantum interference established in the Hilbert space while maintaining the algorithm's memory.
However, experimental realization of quantum speedup for combinatorial optimization is impeded by noise inherent to the current quantum processors~\cite{Harrigan2021_quantum,ebadiQuantumOptimizationMaximum2022,  laydenQuantumenhancedMarkovChain2023,sack2024large, munoz2025scaling,chandarana2025_runtime}.
Since a universal fault-tolerant quantum computer is still far away, great effort has been devoted to the development of quantum algorithms that are resilient to noise.
A paradigmatic example is the quantum approximate optimization algorithm (QAOA)~\cite{farhi2014_quantum, Blekos2024_aReview}, which implements a $Q$-depth quantum circuit containing $2Q$ variational parameters that are optimized through classical feedback loops to find approximate solutions.
Despite its promise, this approach faces practical challenges, particularly the trainability problem~\cite{McClean2018_barren, Cerezo2021_cost, Larocca2025_barren} and the limited solution quality imposed by hardware-constrained circuit depth~\cite{Lykov2023_sampling, Harrigan2021_quantum,pelofske2024scaling}.

Here, we develop a quantum sampling strategy and experimentally demonstrate a hybrid quantum-classical algorithm, named quantum enhanced jumping (Qjump), to address the aforementioned challenges. 
In particular, we propose a new truncated parameter setting scheme highlighted by shallow quantum circuits for the Qjump sampling to guide the search procedure, which only stimulates transitions between promising energy basins while leaving other computing tasks to a classical computer~\cite{Morvan2024_phase, Gao2025_establishing,zhong2021phase}.
We evaluate its performance on a superconducting quantum processor by solving Ising problems of up to $104$ qubits,
with experimental evidence for enhanced solution quality against algorithms such as QAOA and simulated annealing (SA).
We further {envision a quantum hardware} 
with specifications for realizing quantum speedup against SA running on a single-core CPU, which is viable based on available technologies for superconducting qubits.
Our results point out the potential applicability of noisy intermediate-scale quantum (NISQ) processors matched with hybrid quantum-classical algorithms in solving real-world problems.

\begin{figure*}
\centering
\includegraphics[width=1.0\textwidth]{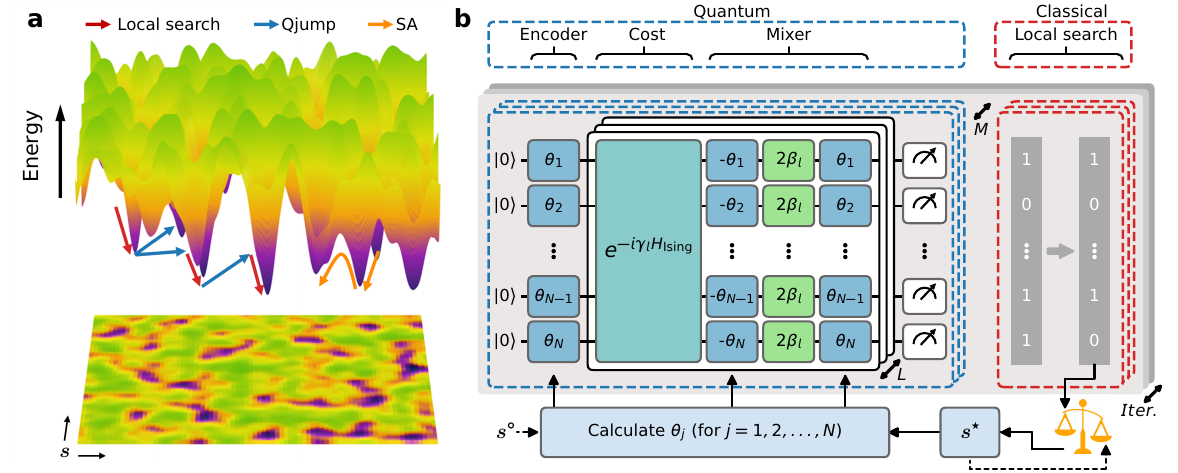}
\caption{
\textbf{Ising model energy landscape and schematic of the Qjump algorithm.}
\textbf{a,} Illustration of a rugged energy landscape of the Ising model in Eq.~\ref{isingH}, with the XY coordinates defined by spin configurations $\bm{s}$. 
Typical optimization trajectories for both the Qjump and SA algorithms are indicated by colored arrows.
\textbf{ b,} Schematic of the Qjump algorithm featuring the quantum-classical workflow. Within the quantum part, the single-qubit gates represented by colored squares of blue (green) in layer $l$ correspond to Bloch sphere rotations around Y-axis (Z-axis), i.e., $\text{Y}(\theta_j)= e^{-i{\theta_j}\sigma^y/{2}}$ $\left(\text{Z}(2\beta_l)= e^{-i\beta_l \sigma^z}\right)$, and $H_{\rm{Ising}}$ in the cost operator is given in Eq.~\ref{isingH}. 
}
\label{fig1}
\end{figure*}

We target on 
{finding the ground state of a general Ising model, a canonical NP-hard problem} with broad practical applications~\cite{Barahona1982,Lucas2014}, which can be formulated by an $N$-qubit Ising Hamiltonian 
\begin{equation}
    \label{isingH}
    H_{\rm{Ising}} = -\sum_{\{j,k\}\in N} J_{jk}\sigma_j^z \sigma_k^z - \sum_{j=1}^N h_j \sigma_j^z,
\end{equation}
where $J_{jk}$ is the coupling strength between qubit pair \{$j$, $k$\}, $\sigma_j^z$ is the Pauli Z operator, and $h_j$ characterizes the local magnetic field. As this Hamiltonian typically refers to a rugged energy landscape spanned by spin configurations with the bitstring notation $\bm{s}$ (e.g., 0101...0110), the optimization goal is to find $\bm{s}$ that approaches the global minimum in energy $E(\bm{s})=\langle\bm{s}|H_\text{Ising}|\bm{s}\rangle$. In this work we consider the scenario of qubits arranged in a square lattice with only nearest-neighbor couplings, which matches the 2D connectivity of our quantum processor, and parameters $J_{jk}$ and $h_j$ take arbitrary real values therein.

We focus on the challenging problem instances as decided by the sets of $J_{jk}$ and $h_j$ values in Eq.~\ref{isingH}, on which classical algorithms such as simulated annealing (SA) struggle but our quantum algorithm might offer an improvement. To determine the challenging instances, $4000$ random instances are first generated, each of which draws $J_{jk}$ and $h_j$ independently from the zero-mean normal distributions with variances of $4$ and $1$, 
respectively, and are then filtered using a Ising-specialized SA solver (SimAn~\cite{isakov2015optimised}) running on a single-core CPU ($2.3$ GHz). 20 problem instances with the highest times to solution (TTS), in the range from $0.02$ to $0.4$ seconds, are selected for the benchmark purpose.

For a rugged energy landscape shown in Fig.~\ref{fig1}\textbf{a}, our proposed Qjump algorithm leverages a quantum sampling circuit which, given an initial bitstring $\bm{s}^\circ$ and by repeating itself multiple (e.g., $M$) rounds, outputs $M$ candidacy bitstrings representing jumps to nearby low-energy basins. Then, a classical local search is performed based on all $M$ bitstrings to return an optimal solution $\bm{s}^\star$ corresponding to the lowest explored energy.
The quantum sampling circuit consists of an encoder layer to generate superpositions, $L$ layers of cost and mixer operators to enable quantum jumps, and a layer of projective measurements to sample one  of the jumps by returning a candidacy bitstring for $N$ qubits. 
As inspired by QAOA, here the cost operator encodes a problem instance of $H_{\rm{Ising}}$ (Eq.~\ref{isingH}) and the mixer operator drives a mixer Hamiltonian non-commuting with $H_{\rm{Ising}}$.
As can be seen in the digital circuit schematic (Fig.~\ref{fig1}\textbf{b}), parameters that need feedback are the rotation angles $\theta_j = 2\arcsin\left( \sqrt{0.5 + (s_j - 0.5)\alpha} \right)$ with the mixing coefficient $\alpha \in [0,1]$, where $s_j$ denotes the $j$-th bit in 
either a given initial input $\bm{s}^\circ$ for a warm start~\cite{egger2021warm} or the best $\bm{s}^\star$ of prior solutions if iterated. $\alpha$ is typically set to an intermediate value, ultimately related to the bit flip ratio of the output $\bm{s}^\star$ compared with the input bitstring. 
Variational parameters used in the cost and mixer operators, i.e., $\gamma_l$ and $\beta_l$ for layer $l$, are predefined using the parameter setting heuristic of a $Q$-depth QAOA ansatz~\cite{basso2022quantum,Shaydulin2023_parameter,Sureshbabu2024_parametersetting} (see Supplementary Section~2C). The entire Qjump algorithm typically cycles the quantum and classical circuit diagrams in Fig.~\ref{fig1}\textbf{b} multiple iterations ($\alpha \equiv 0$ in the first iteration) until a desired solution is obtained. 

\begin{figure*}[ht]
\centering
\includegraphics[width=1.0\textwidth]{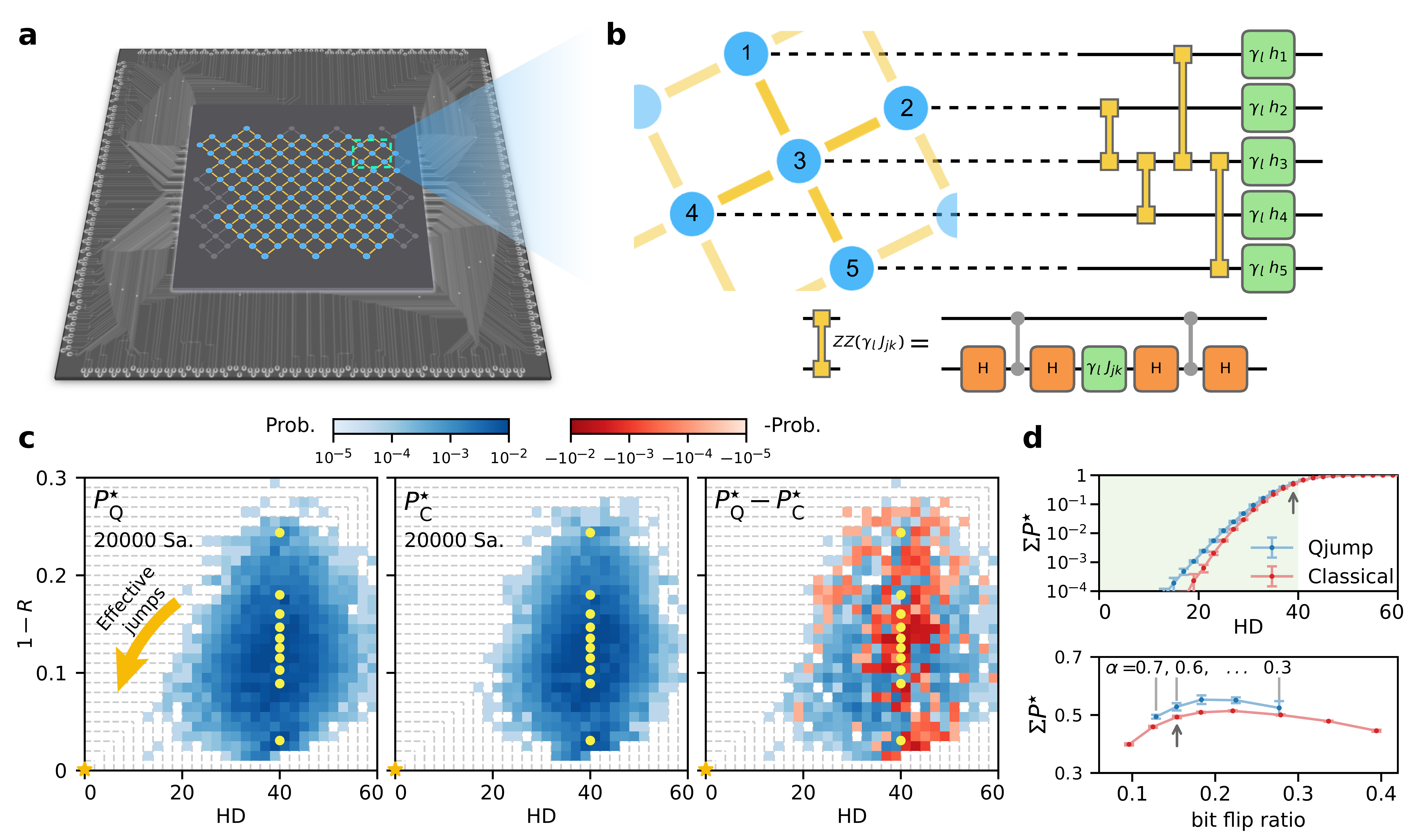}
\caption{
\textbf{104-qubit superconducting quantum processor and performance of the quantum sampler.}
\textbf{a,} Illustration of the superconducting quantum processor, where the $104$ qubits (blue dots) and connecting couplers (yellow lines) used in the experiment are highlighted.
\textbf{b,} Circuit schematic for implementing the cost operator $e^{-i\gamma_l H_{\text{Ising}}}$. 
Here $\text{ZZ}(\gamma_l J_{jk}) = e^{-i{\gamma_l J_{jk}} \sigma^z_j \sigma^z_k/{2}}$, which is compiled into a series of Hadamard gates (orange squares), virtual Z-axis rotations with the angles indicated (green squares), and CZ gates (gray dots connected by lines).
\textbf{c,} Sampling probability $P^{\star}_{\text{Q}}$ (left) as function of energy $1-R$ and Hamming distance HD, by Qjump featuring quantum sampling in comparison with that by classical random sampling $P^{\star}_{\text{C}}$ (middle). Probabilities are obtained by counting bitstrings within the colored unit boxes with sizes of $0.01$ in $1-R$ and $2$ in HD,
{while the total number of bitstrings for normalization is indicated}.
Data here are for problem instance $\#1$ (see Supplementary Section~3A), with $10$ initial $\bm{s}^{\circ}$ all at HD=$40$ (marked by yellow circles). 
Also plotted is the difference between $P^{\star}_{\text{Q}}$ and $P^{\star}_{\text{C}}$ (right).
Since \emph{effective jumps} (illustrated by the thick golden arrow in the $P^{\star}_{\text{Q}}$ panel) refer to those sampling bitstrings that are closer to the global minimum than $\bm{s}^{\circ}$ in terms of both HD and energy values, to identify them we outline a series of square regions all anchoring at the origin by background gray dashed lines, indexed by Hamming distance to the global minimum. 
We note that the aspect ratio of the square regions can be chosen quite flexibly, all yielding similar conclusions in the performance benchmark.
\textbf{d,} Top: Summation of probabilities within the square regions outlined in \textbf{c}, $\sum P^{\star}$, as function of the region index HD. Bottom: $\sum P^{\star}$ within the square region indexed by $\text{HD}=40$, which refers to all \emph{effective jumps} in a single step, 
as function of the bit flip ratio for Qjump with the corresponding $\alpha$ values indicated (blue). Also plotted are the classical random sampling results for comparison (red). Error bars represent standard deviations across five repeated experiments.
}
\label{fig2}
\end{figure*}

The QAOA circuit with a large $Q$ can approximate adiabatic quantum annealing, leading to a high probability of finding the global energy minimum. Our analysis of QAOA circuit dynamics (Supplementary Section~2) indicates that the first few layers primarily facilitate broad-range explorations while the later layers play the function analogous to classical local search. Therefore, we propose to design the Qjump sampling circuit by keeping the first $L$ layers from a $Q$-depth parameter-setting QAOA circuit, named [$L$,$Q$]-sampler. In contrast to QAOA algorithms, the Qjump framework eliminates parameter optimization requirements by leveraging sampling from circuits with fixed transfer parameters according to \cite{basso2022quantum,Shaydulin2023_parameter,Sureshbabu2024_parametersetting} and strategically replaces quantum circuit segments with classical local search routines. 

\begin{figure*}
\centering
\includegraphics[width=1.0\textwidth]{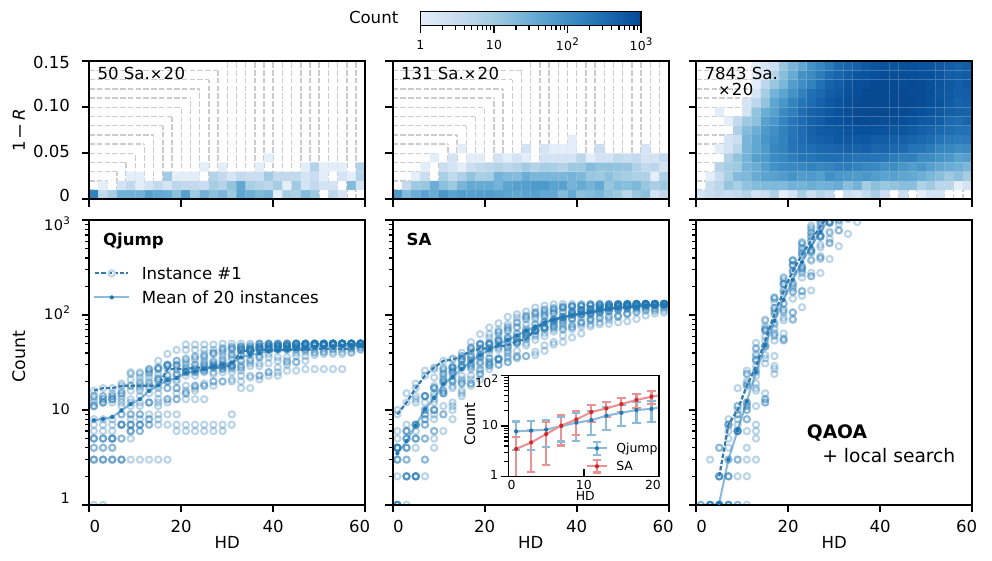}
\caption{
\textbf{Performance of Qjump in comparison with SA and QAOA.}
Solution distributions of Qjump, SA and QAOA summarized over 20 instances. Top row: solution count distribution over 20 problem instances for each algorithm. Bottom row: solution count for individual instances and the means as functions of HD toward the global minimum, which are processed in a manner similar to that in Fig.~\ref{fig2}\textbf{d}. 
Based on the computational speed of the envisioned quantum hardware and classical CPU (See Supplementary Section~3C for the estimated time per run: Qjump $\sim 0.78$~ms; SA $\sim 0.30$~ms; QAOA $\sim 5.1~\mu$s), we first determine the numbers of bitstrings that can be produced by these algorithms in repeated runs within a fixed period of 40~ms (indicated in figure panels for 20 instances), and then run Qjump and QAOA on our superconducting processor to generate the pre-determined numbers of bitstrings for analysis, which take much longer times on our experimental setup.
Mean values extracted from Qjump and SA are shown in the panel inset, with error bars representing the standard deviations over 20 instances. On average, Qjump samples the global minimum 2.23 times more frequently than SA for the 104-qubit Hamiltonians.
}
\label{fig3}
\end{figure*}

We experimentally investigate the Qjump algorithm on a superconducting quantum processor, with 104 frequency-tunable transmon qubits arranged in a two-dimensional square lattice (Fig.~\ref{fig2}\textbf{a}), where controlled $\pi$-phase (CZ) gates can be implemented between any two neighboring qubits with the median gate fidelity around $99.5\%$. All 104 qubits can be simultaneously manipulated by arbitrary rotational gates and jointly measured in the computational basis, as necessary elements in the Qjump diagram (Fig.~\ref{fig1}\textbf{b}), with the median single-qubit gate and measurement fidelities reaching $\sim 99.95\%$ and $> 99\%$, respectively. 
The cost operator $e^{-i\gamma_l H_{\rm{Ising}}}$ in Qjump is compiled into a digital sequence comprising eight layers of CZ gates, with a layer of virtual Z gates (by adding the phase in microwaves for subsequent rotations) and single-qubit rotation gates inserted between two successive CZ layers. Gate twirling is used for CZ gates to mitigate the noise impact\cite{laydenQuantumenhancedMarkovChain2023} (See Supplementary Section 4 for detailed device and circuit information). 

We begin with examining the quantum sampler's performance for a given problem instance, named instance \#1 (see Supplementary Section~3A), and initial guesses $\bm{s}^\circ$, focusing on its sampling probability $P^\star$ toward the global minimum in single-step jumps. 
Here, $P^\star$ represents the probability of a sampled bitstring appearing within a unit area in a two-dimensional plane spanned by the two axes of energy and Hamming distance (HD).
The energy scale is defined as $1-R$ with the approximation ratio $R= E_{\text{exp}} / E_{\text{g}}$, where $E_{\text{exp}}$ is the experimentally reached Ising energy of the sampled bitstring and $E_{\text{g}} < 0$ is the ground-state energy, and HD is determined by the number of flipped qubit sites relative to the global optimal bitstring.
We enumerate 10 initial $\bm{s^\circ}$ that are separated from the global minimum by the same HD (e.g., $40$), and experimentally run the Qjump algorithm with $M=2000$ and one iteration to output $2000$ bitstrings for each $\bm{s^\circ}$, with a total of $20000$ bitstrings to estimate $P^\star$. By defining the unit area as a box with sizes of $0.01$ in $1-R$ and $2$ in HD, scattered plot of $P^\star$ on the energy-versus-HD plane is obtained in Fig.~\ref{fig2}\textbf{c} (left).
Comparing the $20000$ bitstrings with $\bm{s^\circ}$, it is seen that the bit flip ratio essentially conforms to a normal distribution, from which a mean bit flip ratio can be obtained for the quantum sampler.

For predefined parameters necessary to run the Qjump algorithm, we have borrowed the variational parameter values for $\gamma_l$ and $\beta_l$ from the $Q$-depth QAOA ansatz, and tried different $Q$ values up to $20$~\cite{basso2022quantum,Shaydulin2023_parameter,Sureshbabu2024_parametersetting}. We have focused on the shallow-circuit regime by examining layer number $L\in [1,3]$ with $\alpha\in [0.3,0.7]$ for the circuit diagram in Fig.~\ref{fig1}\textbf{b}. 
We find that the Qjump algorithm is quite robust regarding different choices of the predefined parameters, all demonstrating promising capabilities in reaching the vicinity of the global minimum for various problem instances. In the following benchmarks, we choose the [$2$,$20$]-sampler, with a circuit depth of {$33$} including {$1590$} single-qubit and {$728$} two-qubit gates for $104$ qubits, and $\alpha=0.6$, unless otherwise specified, for illustrative purpose only.
{We emphasize that the exact parameter optimization process for Qjump is extremely resource consuming for more than 100 qubits, which is not rigorously attempted here.}

To witness the effectiveness of quantum sampling, we replace the quantum part in Fig.~\ref{fig1}\textbf{b} with classical random sampling, which stochastically flips certain qubit sites on a given $\bm{s}^\circ$ with the flipping ratio equivalent to the mean value for quantum sampling. 
For problem instance $\#1$, scattered plot of $P^\star$ by random sampling under similar conditions is shown in Fig.~\ref{fig2}\textbf{c} (middle) in contrast to the quantum sampler's result, which highlights the fact that Qjump explores more favorably over regions close to the global minimum at the origin [Fig.~\ref{fig2}\textbf{c} (right)]. Since \emph{effective jumps} refer to those sampling bitstrings that are close to the global minimum in terms of both HD and energy values, to better quantify the sampling difference, we define square regions that anchor at the global minimum and shrink down the region sizes (guided by a series of gray dashed-line boundaries and indexed by HD to origin).
The combined probability $\sum P^\star$ within these square regions is shown in Fig.~\ref{fig2}\textbf{d} (top), where larger magnitude of $\sum P^\star$ at small HD values indicates a better performance. Since sampling bitstrings all occurring at $\textrm{HD}< 40$ 
can be considered as effective jumps, we plot in Fig.~\ref{fig2}\textbf{d} (bottom) the effective jumping probabilities at different $\alpha$ values in comparison with the classical sampling with similar bit flip ratios. It is observed that the experimental quantum sampler can perform better than the classical counterpart, even under the experimental imperfections such as non-ideal quantum gates and measurements. This is the case for majority of the problem instances.

Now we move on to quantifying the solution quality of the Qjump algorithm with a quantum-classical hybrid workflow, in comparison with the popular QAOA and SA algorithms based on 20 problem instances of the $104$-qubit Ising model. We examine the chances of the solutions (bitstrings) that approach close to the global minimum, under the prerequisite that our comparison must take into account the different algorithmic runtimes. Returning one optimal bitstring per run, these algorithms can yield quite different numbers of bitstrings within the same period due to different algorithmic structures and parameter settings. For a fair comparison, instead of plotting sample probabilities as done in Fig.~\ref{fig2}, here and below we examine the distribution of sample occurrences, i.e. counts of bitstrings in unit boxes without normalization, in the energy-HD plane. Larger counts close to the global minimum certify a better solution quality.

Our experimental setup cannot run algorithms fast enough to yield a significantly large number of bitstrings within a short period to win the competition (See Supplementary Section 3C for details). Instead, we look into the possibility of assembling such a state-of-the-art quantum hardware based on available technologies for superconducting qubits, and wish to testify that this quantum hardware may demonstrate quantum speedup by running the Qjump algorithm in contrast to SA running on a single-core CPU.
For this purpose, we make a couple of assumptions of the quantum hardware: Similar to the requirements for quantum error correction, the envisioned quantum hardware is able to perform a sequence of actions including qubit reset, gate operation, measurement, classical local search and comparison, and feedback control to implement gate set for the next iteration, all controlled by FPGA with high programming flexibility and without resorting to any external classical computer; according to literature-reported speed of quantum operations and measurements, the quantum sampler can reach a sampling frequency of $0.4$~MHz~\cite{acharyaSuppressingQuantumErrors2022,salatheLowLatencyDigitalSignal2018,ai2024quantum}, and FPGA can process classical information equally fast in comparison to the CPU that runs the SA algorithm {(see Supplementary Section~3C for the workflow of Qjump)}.

With aforementioned assumptions and the knowledge of the actual runtimes of classical operations, we first determine the number of bitstrings that could be produced by each algorithm within a fixed period of 40 ms.
We then repeatedly run Qjump with $12$ iterations and QAOA with $Q=6$ on our superconducting processor, and SA with $700$ sweeps on a single-core CPU to generate these pre-determined numbers of bitstrings for analysis.  We append classical local search to QAOA, which is necessary for the algorithm to truly locate the global minimum. QAOA with $Q=6$ has a circuit depth of $97$ layers with roughly $4500$ single-qubit gates and $2100$ CZ gates. 
The optimal cycling parameter for each algorithm, i.e., number of iterations (Qjump), number of sweeps (SA), or depth of circuits (QAOA), is chosen where TTS reaches close to the best (see next).

The resulting distributions of sample occurrences in the energy-HD plane averaged over the 20 problem instances by Qjump, SA, and QAOA are displayed back to back in the top row of Fig.~\ref{fig3}. Although Qjump yields much less ($50$ for one instance) bitstrings in 40~ms, it appears to explore close to the global minimum with more weights. To better quantify this difference, we again plot sample occurrences for the 20 instance in the bottom row of Fig.~\ref{fig3} as functions of square region index HD toward the origin. It is seen that averaged sample occurrence by Qjump is highest as HD goes to zero, certifying its best performance in reaching the global minimum.

\begin{figure}[t]
\centering
\includegraphics[width=1.0\linewidth]{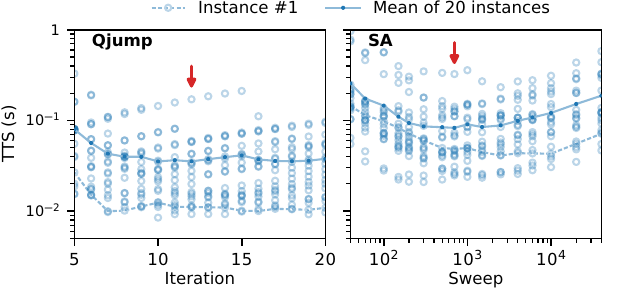}
\caption{
\textbf{Time-to-solution (TTS) estimation for Qjump and SA.}
Each empty circle represents the TTS value for a certain problem instance obtained by running the Qjump (SA) algorithm with the specified iteration (sweep) number. Dots connected by lines indicate the mean for the 20 problem instances, and the arrow points to the optimal iteration/sweep number used for the algorithmic benchmark in Fig.~\ref{fig3}. According to the minimal TTS values, on average Qjump on the envisioned quantum hardware outperforms SA on a single-core CPU by a factor of $2.34$ for the $104$-qubit Ising model (see Supplementary Section~4D for cases with fewer qubits). TTS values for instance $\#1$ are highlighted by dashed lines. 
The QAOA algorithm appended by classical local search is almost ineffective in reaching the global minimum, whose TTS values, estimated to be more than $4$~s, are not shown here.
}
\label{fig4}
\end{figure}

Now we discuss the speedup potential of Qjump, based on the aforementioned assumptions for the envisioned quantum hardware and the knowledge of the actual runtimes of classical operations.
We quote the standard metric $\text{TTS} = t_\text{r}\log(1-0.99)/ \log(1-p_{\text{s}})$, which specifies the time required to reach the global minimum with $99\%$ probability, where $t_\text{r}$ ($p_{\text{s}}$) is the time (success probability of sampling the global minimum) per run of the algorithm. Figure~\ref{fig4} shows the TTS estimation for Qjump and SA (data for QAOA not shown), based on which we select the optimal cycling parameter for each algorithm. A more important message is that, with available technologies for superconducting qubits, Qjump running on the envisioned quantum hardware might outperform SA on a single-core CPU by a factor of 2.34 with respect to the TTS metric, which is as expected from Fig.~\ref{fig3}. Of course, the performance of classical heuristics can be significantly improved by utilizing parallel computing techniques based on GPU, which is not considered in this topic.

Nevertheless, the possible advantage compared with single-core CPU is still stimulating, which potentially leads to many open and interesting questions if the benchmark can be extended to larger scales.
On one hand, the scaling performance of the approximation optimization heuristics is known to be unstable at small system sizes~\cite{Ronnow2014_defining}. On the other hand,
due to the limited size of the current experiment, the time complexity of Qjump is not yet understood, and the eventual answer heavily relies on progresses in scaling up the experimental demonstration.
Meanwhile, influence of gate and measurement infidelities on the shallow-circuit Qjump sampler remains unclear, and it would be interesting to explore how to rigorously optimize the quantum sampler regarding both the parameter settings and circuit structure. 

As we demonstrated in our work, the hybrid workflow of the Qjump that judiciously combines most efficient aspects of both classical and quantum subsystems, has prospect of gaining advantage against previously known algorithms. Namely, pre-trained parameter setting scheme avoids variational updates and thus mitigates the risk of encountering a barren plateau. Circuit truncation and use of steepest descent local search enhances the algorithm’s robustness against noise by reducing the circuit depth and offloading the exploration of subtle landscape features to classical subsystem. At the same time, we demonstrated that sampling from shallow quantum circuits with novel use of truncated parameter setting can serve as an efficient tool for broader landscape exploration and escape from local minima, which is the most challenging part of classical optimization.
This result can be linked to previous theoretical works on quantum advantage achieved with shallow quantum circuits \cite{Bravyi2018, farhi2019_quantum, Bravyi2020}, although we do not claim here quantum advantage in strong sense.

Here, we have only considered Ising problems with nearest-neighbor connections. Higher connectivity cases can be converted to nearest-neighbor ones at the cost of additional swapping gates, with the number of gates scaling polynomially with the qubit number~\cite{hashim2022optimized}. Also, practical optimization problems typically involve thousands to millions of variables which may require much more than the 104 qubit as our experiment demonstrated. It is certainly interesting to run the Qjump algorithm on thousands of qubits or more with further improved qubit metrics, which might lead to the ultimate goal of quantum advantage on NISQ devices, but our current demonstration with 104-qubit quantum circuit can be treated as a subsystem which might still offer some jumping ability~\cite{ferguson2025quantum}. Additionally, as escaping local minima is needed in many optimization algorithms, Qjump can be combined with traditional tabu search or Metropolis method, or used as a heuristic module in exact algorithms, e.g., the branch and bound algorithm.
In practice, different heuristics, be it quantum or classical, can show large performance deviations in solving different problem instances, which necessitates algorithm selection~\cite{Moussa_2020}. 
In this perspective, the Qjump algorithm presents a promising alternative in the heuristic ensemble. 

\vspace{.5cm}
\noindent\textbf{Acknowledgement}\\
{The device was fabricated at the Micro-Nano Fabrication Center of Zhejiang University.
We acknowledge the support from the Innovation Program for Quantum Science and Technology (Grant No. 2021ZD0300200), the National Natural Science Foundation of China (Grant Nos. 92365301, 12274368, 12274367, 12322414, 12404570, 12404574), and the Zhejiang Provincial Natural Science Foundation of China under Grant (Grant Nos. LR24A040002 and LDQ23A040001). H. Wang is also supported by the New Cornerstone Science Foundation through the XPLORER PRIZE.
}

\vspace{.5cm}
\noindent\textbf{Author contributions}\\
X.Z. and F.J. carried out the experiments and analyzed the experimental data under the supervision of H. Wang; Z. Zou, P.M., and M.L. conducted the theoretical analysis supervised by M.-H.Y.; H.L. and J.C. fabricated the device supervised by H. Wang; H. Wang, X.Z., Z. Zou, P.M., F.J., and M.-H.Y. co-wrote the manuscript; 
H. Wang, Q.G., Z.W., C.S, P.Z., X.Z., F.J., Y. Wu, C.Z., Y.G., N.W., Y.Z., A.Z., F.S., Z.B., Z. Zhu, J.Z., Z.C., Y. Han, Y. He, Han Wang, J.-N.Y., Y. Wang, J.S., G.L., Z.S., J.D., and H.D. contributed to experimental setup. All authors contributed to the discussions of the results.

\clearpage
\bibliographystyle{naturemag}
\bibliography{qjump}

\end{document}


\nolinenumbers
\title{Supplementary Information for\\
Combinatorial optimization enhanced by shallow quantum circuits with 104 superconducting qubits
}
\maketitle

\tableofcontents
\beginsupplement

\section{A brief overview of combinatorial optimization algorithms} 

Despite the rapid development of hardware platforms, the Noisy Intermediate-Scale Quantum (NISQ) era has yet to produce a truly transformative application. This paper introduces a quantum-classical hybrid optimization algorithm for solving combinatorial optimization problems, providing a promising avenue for achieving quantum enhancement.

Many combinatorial optimization problems, such as the Traveling Salesman Problem and the Knapsack Problem, are NP-hard. These problems are ubiquitous in real-world scenarios, including logistics, scheduling, and resource allocation, and require an exhaustive search through an exponentially large space to find the optimal solution. 
Traditional algorithms for these problems are broadly categorized into two types: exact methods and heuristic methods.
Exact methods, such as integer optimization techniques used by commercial solvers like Gurobi, guarantee the accuracy of the solution but are often computationally expensive.
Heuristic methods, inspired by physical phenomena such as simulated annealing (SA) \cite{Kirkpatrick1983}, tabu search~\cite{glover1989tabu} and genetic algorithms~\cite{holland1992genetic}
iteratively explore the variable space to minimize the objective function. 

In this work, we focus on solving the Ising model, as most combinatorial optimization problems can be mapped to its ground state problem \cite{Lucas2014_Ising}. 
While classical heuristic algorithms can be accelerated using caching techniques to reduce computation time, their exploratory capabilities remain fundamentally limited. Their search steps are fixed and universal, making them less effective for specific problem structures. In contrast, quantum optimization algorithms, such as the Quantum Approximate Optimization Algorithm (QAOA) \cite{farhi2014_quantum} and its variants~\cite{Blekos_2024}, the Quantum Imaginary Time Evolution (QITE) algorithm~\cite{mcardle2019variational,Wang_2025}, quantum walks~\cite{marsh2020combinatorial}, and quantum annealing~\cite{kadowaki1998quantum}, leverage problem-specific information to perform a global search across the quantum state space. By quantum tunneling and quantum interference, these algorithms have the potential to outperform classical methods, as demonstrated by studies~\cite{munoz2025scaling,basso2022quantum,montanezbarrera2024universalqaoaprotocolevidence}.

One significant challenge for quantum optimization algorithms is the excessive circuit depth required to achieve a quantum advantage. Even the least resource-intensive QAOA requires a considerable number of qubits and layers (e.g., 12 layers for hundreds of qubits, as reported in Ref.~\cite{basso2022quantum}), which exceeds the capabilities of current hardware~\cite{pelofske2024short,harrigan2021quantum}. To address this, we propose a hybrid quantum-classical algorithm, named quantum enhanced jumping (Qjump). By alternating between classical and quantum processors, the algorithm delegates the exploration of the search space to the quantum processor and the refinement of the solution to the classical processor. This hybrid approach can potentially achieve a quantum speedup even with current NISQ hardware.

\section{Theoretical analysis}  
    The excellent performance of the Qjump algorithm stems from the use of shallow-depth Warm-started QAOA\cite{Egger_2021,Tate_2023} circuits with novel truncated parameter setting technique, which
    can enhance the probability of landing in a better energy basin through constructive interference. 
    This contrasts with the classical heuristics that escape local energy basins through random exploration, leading to an equal probability of landing on both poor and good bitstrings. In the following section, we will explain how the quantum circuit achieves this.

\subsection{Principle of QAOA}

The standard QAOA starts from an equal superposition state $\ket{+}^{\bigotimes N}=\frac{1}{\sqrt{2^N}}\sum_{y=1}^{2^N}\ket{\bm{s}^y}$, where $\bm{s}^y$ represents all possible bitstrings in the Hilbert space. The algorithm then executes a quantum circuit consisting of alternating cost and mixer layers. For the simplest case of a single-layer QAOA circuit, with $\gamma$ and $\beta$ as the parameters for the cost and mixer layers, respectively, the probability amplitude for obtaining a specific bitstring $\bm{s}^x$ is given by \cite{Diez-Valle2024}:
\begin{equation}
    F_x = \frac{1}{\sqrt{2^N}} \sum_y \left\langle \bm{s}^x \left| e^{-i\beta H_{\text{M}}} e^{-i\gamma H_{\text{Ising}}} \right| \bm{s}^y \right\rangle = \frac{1}{\sqrt{2^N}} \sum_y \left\langle \bm{s}^x \left| e^{-i\beta H_M} \right| \bm{s}^y \right\rangle e^{-i\gamma E_y},
\end{equation}
where $H_{\text{M}} = \sum_{j=0}^N X_j $ and $H_{\text{Ising}}$ are the mixer and Ising problem Hamiltonians, respectively. $X_j$ represents a $\pi$ rotation around the X-axis of the Bloch sphere for qubit $j$, and $E_y$ is the Ising energy for $\bm{s}^y$.  
After cost layer, energies are imprinted onto the phase of the state. Since phase information cannot be directly measured, the mixer layer is used to amplify the amplitudes of the optimal solutions.
The inner product can be expressed as $ \left\langle \bm{s}^x \left| e^{-i\beta H_{\text{M}}} \right| \bm{s}^y \right\rangle = \cos^{N-d_{xy}}(\beta) (-i\sin(\beta))^{d_{xy}} $, where $ d_{xy} $ is the Hamming distance between $ \bm{s}^x $ and $ \bm{s}^y $. Thus, $ F_x $ can be rewritten as:
\begin{equation}\label{eq:F_x}
   F_x = \frac{1}{\sqrt{2^N}} [\cos(\beta)]^N \cdot \sum_y e^{-i(\gamma E_y + \frac{\pi}{2} d_{xy})} [\tan(\beta)]^{d_{xy}}. 
\end{equation}
The sum in this formula can be visualized as the superposition of vectors on complex plane, each corresponding to a bitstring $ \bm{s}^y $. The phase of each vector is determined by its energy and Hamming distance to $\bm{s}^x$ via expression $\gamma E_y + \frac{\pi}{2} d_{xy}$, while its length is proportional to $[\tan(\beta)]^{d_{xy}}$. 

\begin{figure*}
\centering
\includegraphics[width=0.95\textwidth]{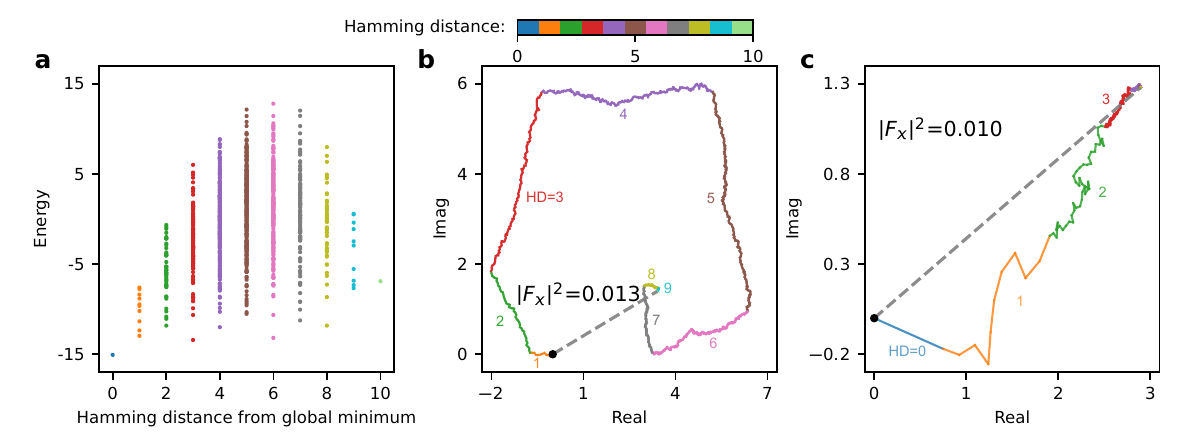}
\caption{
\textbf{The energy-Hamming-distance correlation for the QAOA circuit.}
\textbf{a.} Energy distribution as a function of Hamming distance from the global minimum.
\textbf{b.} Complex plane  plot of each component of $F_x$ for the $Q=1$ QAOA circuit ($\beta\approx0.67$), ordered by Hamming distance.
\textbf{c.} Complex plane  plot of each component of $F_x$ for the final layer of the $Q=3$ QAOA circuit ($\beta\approx0.23$), ordered by Hamming distance.
}
\label{fig:qaoa_paths}
\end{figure*}

We demonstrate this visualization using a 10-qubit regular-4 Ising model. Since our primary interest is the probability of finding the global minimum, we choose the global optimal bitstring as $\bm{s}^x$ and plot the distribution of all bitstrings' energies and their Hamming distances from $\bm{s}^x$ in Fig.~\ref{fig:qaoa_paths}\textbf{a}. 
To visualize the superposition, we group the vectors by their Hamming distance, and then sort the vectors by their corresponding energies within each distance group. Sequentially connecting these vectors forms an open-ended, meandering curve. The distance between the start and the open end of the curve corresponds to the probability amplitude of obtaining $\bm{s}^x$.

\begin{figure*}
    \centering
    \includegraphics[width=0.8\textwidth]{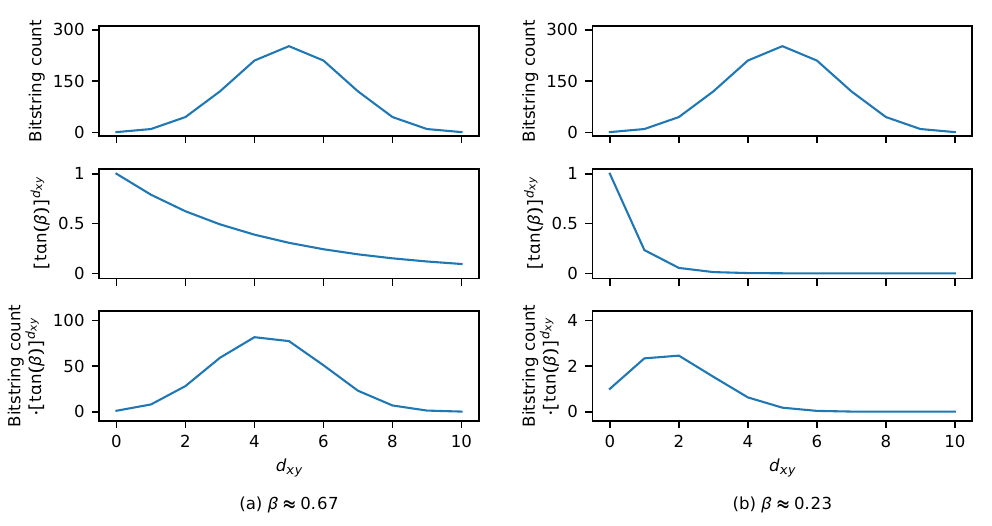}
    \caption{
    \textbf{The contributions from different Hamming distance.}
    Panels (a) and (b) illustrate the contribution of bitstrings at varying Hamming distances to $F_x$ for $\beta\approx0.67$ and $\beta\approx0.23$. Each panel shows, from top to bottom, the count of bitstrings at each Hamming distance, the corresponding $[\tan(\beta)]^{d_{xy}}$ and their product.
    }
    \label{fig:contributions_beta}
\end{figure*}

As shown in Fig.~\ref{fig:qaoa_paths}\textbf{b} and \textbf{c}, we display the results for a $Q=1$ QAOA circuit (with $\beta\approx0.67$) and the final layer of a $Q=3$ QAOA circuit (with $\beta\approx0.23$). For the $Q=1$ circuit, bitstrings with Hamming distances ranging from 2 to 8 contribute significantly to the overall path. In contrast, for the final layer of the $Q=3$ circuit, only bitstrings with $d_{xy}\leq 3$ have a significant contribution.
This effect is further illustrated in Fig.~\ref{fig:contributions_beta}, where we plot the total contributions from bitstrings at different Hamming distances for these two cases. The comparison of these distributions reveals that as $\beta$ decreases, the contribution from bitstrings with a small Hamming distance becomes significantly more pronounced, despite the exponential growth in the number of bitstrings at larger distances.
Note that all numerical simulations of quantum circuit dynamics and sampling were performed using MindQuantum v0.10 software \cite{xu2024mindspore}.

It should be noted that while these results are for a single QAOA layer, the same principles apply to deeper circuits. The only difference is that the initial superposition state for each layer changes, while the cost and mixer layers themselves function in the same manner. 
In a sufficiently deep QAOA circuit, the annealing schedule involves a gradual increase in $\gamma$ and a progressive decrease in $\beta$ with increasing depth. This behavior effectively causes the later stages of the QAOA circuit to act essentially similar to a classical local search, as the contributions to $F_x$ are dominated by bitstrings within a very narrow Hamming distance.

\begin{figure*}[ht]
    \centering
    \includegraphics[width=0.8\textwidth]{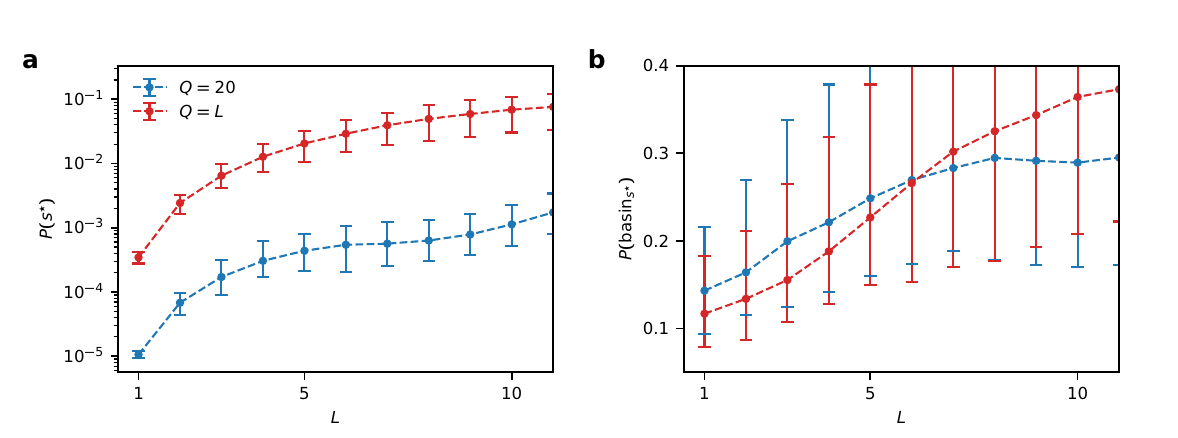}
    \caption{\textbf{Numerical simulations of $L$-layer QAOA circuits with ansatzes of $Q=20$ and $Q=L$.} The results are averaged over 100 random 20-qubit regular-4 Ising instances with gaussian weights and external fields. The error bars represent the lower and upper quartiles.
    \textbf{a, } The probability of obtaining the exact global optimal solution $s^\star$. 
    \textbf{b, } The probability of reaching global optimal basin, where $Q=20$ have better performance with shallow circuits.}
    \label{fig:DQA_depth}
\end{figure*}

To validate the effectiveness of these multi-layer circuits on problems that are challenging for classical local search, we simulate QAOA circuits with varying layers ($L$) on 100 random instances of the 20-qubit regular-4 Ising model. For this simulation, we consider two different QAOA ansatz: $Q=L$ and $Q=20$.
We define the energy basin of a local minima as the set of all bitstrings that can reach that local minima after employing a steepest descent local search.
As shown in Fig.~\ref{fig:DQA_depth}, in the shallow circuit region for the $Q=20$ ansatz, the probability of reaching the exact global minimum is smaller than for the $Q=L$ ansatz. However, the probability of reaching the global optimal energy basin is higher. This shows that when used in a hybrid approach with a classical local search with shallow circuits, the $Q=20$ ansatz produces higher-quality solutions compared to the $Q=L$ ansatz.

For the analysis of large-scale systems, where it is computationally infeasible to calculate the energy of every bitstring or to simulate the quantum circuit, we rely on a mathematical formalization. We adopt the methodology proposed in references~\cite{Diez-Valle2023, Diez-Valle2024}, which assumes that the joint distribution of energy and Hamming distance follows a bivariate Gaussian distribution. Specifically, we model the energy distribution as $E_y\sim \mathcal{N}(0,\sigma^2_E)$, the Hamming distance distribution as $d_{xy}\sim{\mathcal{N}}(N/2,N/4)$, assume a non-zero covariance between them, i.e., $\text{Cov}(E_y,d_{xy})\neq 0$. 
Furthermore, we consider how the distribution of $d_{xy}$ is affected by the $[\tan(\beta)]^{d_{xy}}$ weighting factor. This weighting term creates a new, biased distribution for the Hamming distance labeled $\tilde{d}_{xy}$.
Based on these assumptions, the probability of obtaining a $\bm{s}^x$ can be approximated as:
\begin{equation}
P(\bm{s}^x ) = |F_x|^2 \sim e^{-\gamma^2 \sigma_{E}^2 - \pi \gamma \text{Cov}(E_y, \tilde{d}_{xy})},
\label{eq:Fx}
\end{equation}
The probability peaks at $\gamma^{\star}=-\frac{\pi\text{Cov}(E_y, \tilde{d}_{xy})}{2\sigma_{E}^2}$ with the maximal value of $P^{\star}(\bm{s}^x )\sim e^{\pi^2\text{Cov}(E_y, \tilde{d}_{xy})/4\sigma_{E}^2}$.
For a given $ \sigma_{E} $, a stronger energy-Hamming-distance correlation, $\text{Cov}(E_y, \tilde{d}_{xy})$, results in a higher probability of obtaining $\bm{s}^x$. These results will be used in the next section to analyze the circuit's behavior.

\subsection{Principle of warm-start QAOA}

A major limitation of standard QAOA is the restriction on achievable circuit depths due to current hardware constraints. In Qjump, we address this using the technique of Warm-start QAOA (WS-QAOA)\cite{Egger_2021,Tate_2023}, which leverages information from a previous iteration to initialize the quantum circuit. This approach focuses the quantum sampling on finding an improved solution within the vicinity of a known solution.
Given an initial solution $\bm{s}^{\circ}$, the amplitude $F_{x}$ after a single WS-QAOA layer is given by:
\begin{equation}
   F_{x} =\braket{\bm{s}^x|\bm{s}^{\circ},\theta,\gamma,\beta}  =  [\cos(\theta/2)]^N \cdot (\cos{\beta}+i\sin{\beta}\cos{\theta})^N \cdot \sum_{y} e^{-i\gamma E_{y}} [\tan(\theta/2)]^{d_{\circ y}}\left (\frac{i\sin{\theta}\sin{\beta}}{\cos{\beta}+i\sin{\beta}\cos{\theta}}\right)^{d_{xy}},
\label{Eq:wsqaoa}
\end{equation}
where $\theta = 2\arcsin\left( \sqrt{0.5 - 0.5\alpha} \right) $, and $\gamma$ and $\beta$ are the circuit parameters of the first layer.
This formulation introduces an exponential suppression factor, $[\tan(\theta/2)]^{d_{\circ y}}$,  which reshapes the initial uniform superposition into a distribution concentrated around $\bm{s}^{\circ}$. 

In comparison with standard QAOA, warm-start QAOA localizes the search space around the initial solution $\bm{s}^{\circ}$. When an proper $\bm{s}^{\circ}$ is chosen near the global minimum $\bm{s}^{\star}$, the basin containing $\bm{s}^\star$ is favored by the high energy-Hamming-distance correlation, leading to a higher sampling probability. Conversely, bitstrings far from $\bm{s}^{\star}$ are negatively affected by this correlation. 
Therefore, introducing a better initial guess such as the Qjump does for later iterations is helpful.

\begin{figure*}[!ht]
\centering
\includegraphics[width=\textwidth]{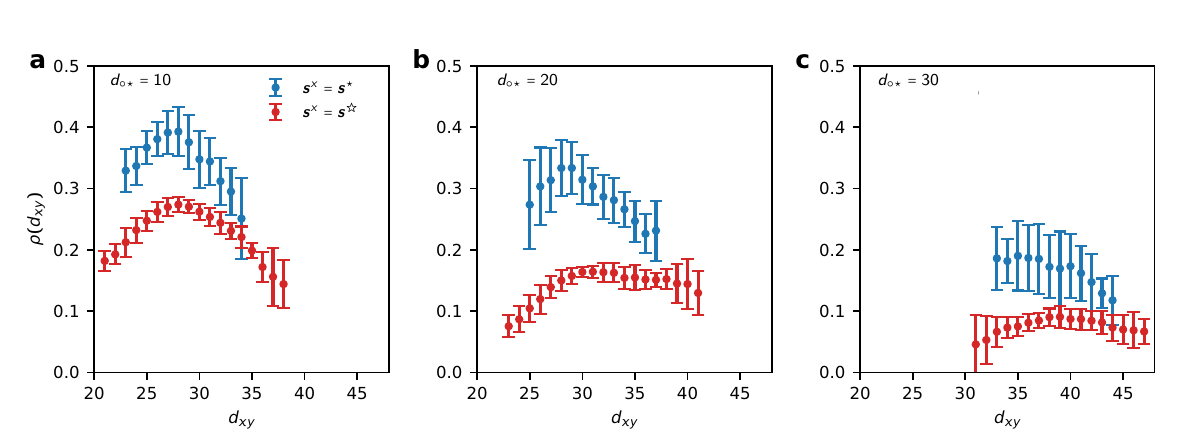}
    \caption{
    \textbf{The Energy-Hamming-distance correlations of bitstrings $\bm{s}^y$ sampled from the distribution $p(\bm{s}^y|\bm{s}^x,\bm{s}^{\circ})$}. The plots show the local covariance when starting from local minima ($\bm{s}^{\circ}$) at various Hamming distances from the global minimum ($\bm{s}^{\star}$). Error bars represent the standard deviation of bitstrings sampled at each $d_{xy}$.
    \textbf{a, } $\bm{s}^{\circ}$ with a Hamming distance of 10 from $\bm{s}^{\star}$. 
    \textbf{b, } $\bm{s}^{\circ}$ with a Hamming distance of 20 from $\bm{s}^{\star}$.
    \textbf{c, } $\bm{s}^{\circ}$ with a Hamming distance of 30 from $\bm{s}^{\star}$.
    }
\label{suppfig:Cov_104q}
\end{figure*}

To demonstrate the distinct behavior of quantum jumping in contrast to its classical counterpart, we performed a numerical simulation on instance \#1 from the main text ($N=104$) with $\alpha=0.6$ and $\gamma$, $\beta$ from the first layer of a $Q=20$ QAOA ansatz. We start from $\bm{s}^\circ$ trapped in local minima (e.g., at a Hamming distance of 10 from $\bm{s}^{\star}$). Using a Monte Carlo method and an approximation similar to the one introduced in Section~2A, we sample bitstrings from the conditional distribution $p(\bm{s}^y|\bm{s}^x,\bm{s}^{\circ}) \sim [\tan(\theta/2)]^{d_{\circ y}}\left |(i\sin{\theta}\sin{\beta})/(\cos{\beta}+i\sin{\beta}\cos{\theta})\right|^{d_{xy}} $. We define two target states: the global minimum $\bm{s}^x = \bm{s}^\star$, and an opposite distant state $\bm{s}^x = \bm{s}^{\medstar}$  (e.g. at a Hamming distance 10 from $\bm{s}^\circ$ and 20 from $\bm{s}^\star$). Due to the computational intractability of calculating the full covariance over all possible bitstrings, we instead sample 10000 bitstrings for each target state and calculate the local covariance $\rho(d_{xy})$ between the energy $E_y$ and the Hamming distance $d_{xy}$ within a small range (e.g., Hamming distance from $\bm{s}^y$ within 4). This quantity $\rho(d_{xy})$  correlates with the probability of obtaining the target state $\bm{s}^x$, with a higher value indicating a higher probability. As shown in Fig.~\ref{suppfig:Cov_104q}, the results for each target state are averaged over 10 different $\bm{s}^\circ$ and 10 different $\bm{s}^x$, respectively.

In classical jumping, starting from $\bm{s}^\circ$, the states $\bm{s}^\star$ and $\bm{s}^\medstar $ have the same Hamming distance and thus an equal probability of being reached. In contrast, quantum jumping simulation shows a very different behavior. As shown in Fig.~\ref{suppfig:Cov_104q}, the energy-Hamming-distance correlations for the target states are clearly different. The global minimum exhibits a much better correlation than the distant ones, which explains why quantum jumping is superior to classical jumping in guiding the search towards the optimal region.

\subsection{Parameter setting scheme}
QAOA is a variational algorithm that employs alternating cost and mixer layers, with $\gamma_{l}$ ($\beta_{l}$) donate the parameter for the $l$-th cost (mixer) layer. Traditionally, these parameters are optimized through classical optimizer, with the loss function evaluated on a quantum processor. Recent work by Shree Hari Sureshbabu {\it et al.}~\cite{Sureshbabu2024_parametersetting} established near-optimal QAOA parameters for arbitrary weighted Ising model, by transferring parameters classically derived from infinite size unweighted Ising problems. We utilize these near-optimal parameters to construct our quantum sampler.

    \indent The procedure to construct a [$L$,$Q$]-sampler is as follows: 
    
    \begin{enumerate}
    \item The QAOA parameters $\gamma^{\rm inf}_l$, $\beta^{\rm inf}_l$ (for the $l$-th layer $l=1,\dots,Q$) optimized for large-girth regular unweighted graphs in the infinite-size limit are prepared in advance~\cite{basso2022quantum}. We then select the parameters of the first $L$ layers from this set.
    
    \item Given the Ising problem with $J_{jk}$ and $h_j$, calculate the rescale factor $A$ to account for weight distribution of the Ising model, as suggested in works \cite{Sureshbabu2024_parametersetting}:
   
\begin{equation}
	A = \sqrt{\frac{1}{C_J}\sum_{\{j,k\} \in N} J_{jk}^2 + \frac{1}{C_h}\sum_{j=1}^N h_j^2},
\end{equation}   
    
 where $C_J$ and $C_h$ are the number of non-zero elements in $J_{jk}$ and $h_j$.

    \item Calculate the average degree of the graph $D$:
    \begin{equation}
        D = \frac{C_J}{N}.
    \end{equation}

    \item The final QAOA parameters are given by:
    \begin{equation}
        \gamma_{l} = A  \arctan{\frac{1}{\sqrt{D-1}}}\times\gamma^{\rm inf}_{l},\quad \beta_{l} = \beta^{\rm inf}_{l}.
    \end{equation}
    
\end{enumerate}

\section{Numerical simulations}

This section outlines the numerical simulation details for the experiment, including problem instance generation, the simulated annealing algorithm employed, and time analysis of the various components within Qjump. 

\subsection{Problem instances}
All instances in this experiment are 2D Ising models with external fields, and their connectivity is shown in main text Fig.~2\text{a}. To align with the chip's topology, we used a rotated lattice, which maximizes the utilization of the qubits. The number of qubits in the lattice scales as $2 \times L \times (L+1)$ with increasing edge lengths.
For better illustration, we selected instances with 60, 84, and 104 qubits, corresponding to $L=5,6$, and $\sim7$. 
We focus on the following Ising model:
\begin{equation}
    \label{isingH}
    H_{\rm{Ising}} = -\sum_{\{j,k\}\in N} J_{jk}\sigma_j^z \sigma_k^z - \sum_{j=1}^N h_j \sigma_j^z,\quad J_{jk} \sim \mathcal{N}(0,4),\quad h_{j} \sim \mathcal{N}(0,1),
\end{equation}
where qubits are arranged in a square lattice with only nearest-neighbor couplings, matching the 2D connectivity of our quantum processor.

In the field of optimization, the no-free-lunch theorem states that any two classical optimization algorithms should, on average, perform equally well across all possible optimization instances. To better study the performance of quantum optimization algorithms and their differences from classical algorithms, we filtered the benchmark instances.  Given that practical optimization often involves running multiple algorithms in parallel and selecting the best result, our interest lies less in instances where classical algorithms like SA perform well. Instead, we focus on challenging instances where classical algorithms struggle and where quantum computing might offer an improvement.

\begin{figure*}
    \centering
    \includegraphics[width=1.0\textwidth]{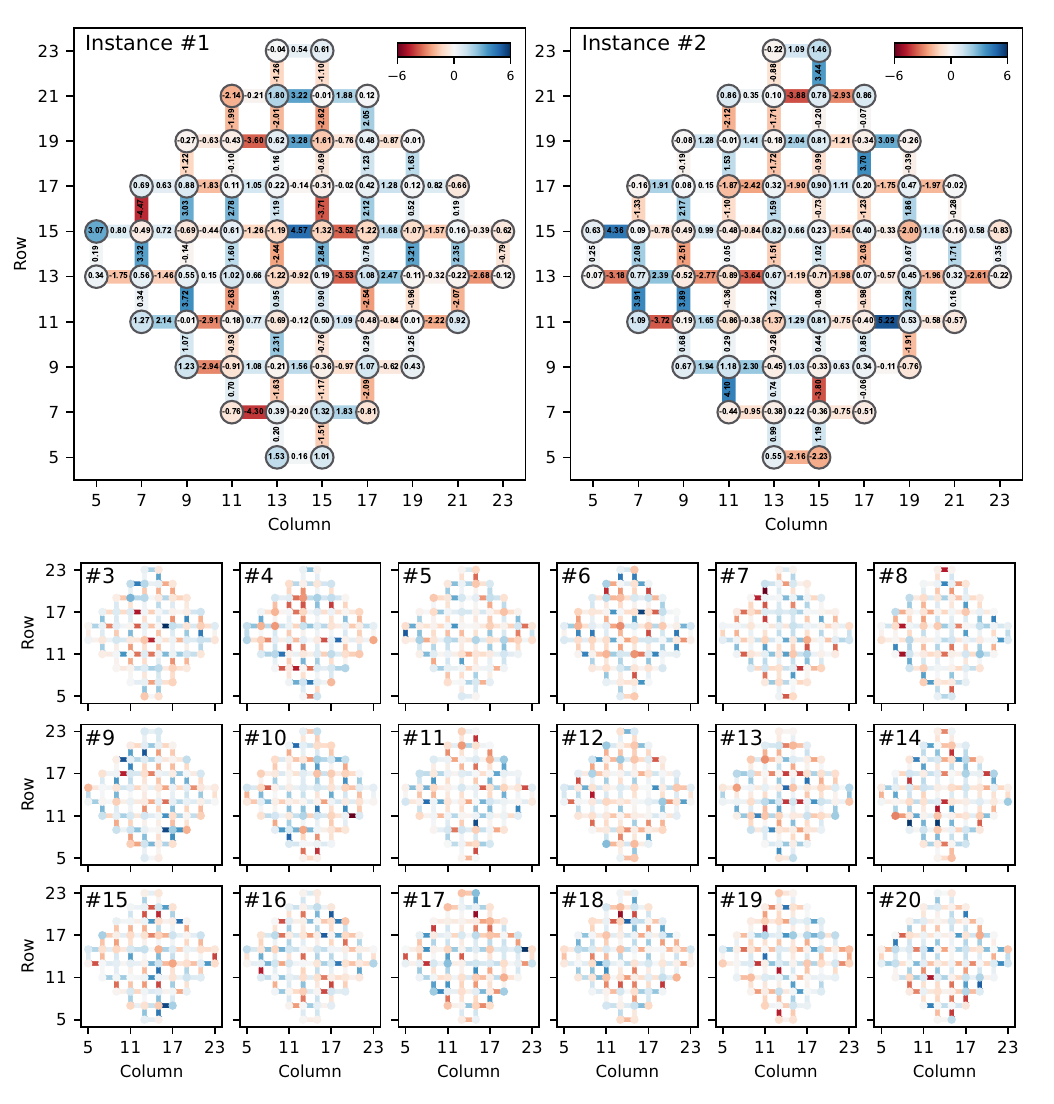}
    \caption{
    \textbf{Problem instances for $N=60$.}
    }
    \label{fig:inst_60q}
\end{figure*}

\begin{figure*}
    \centering
    \includegraphics[width=1.0\textwidth]{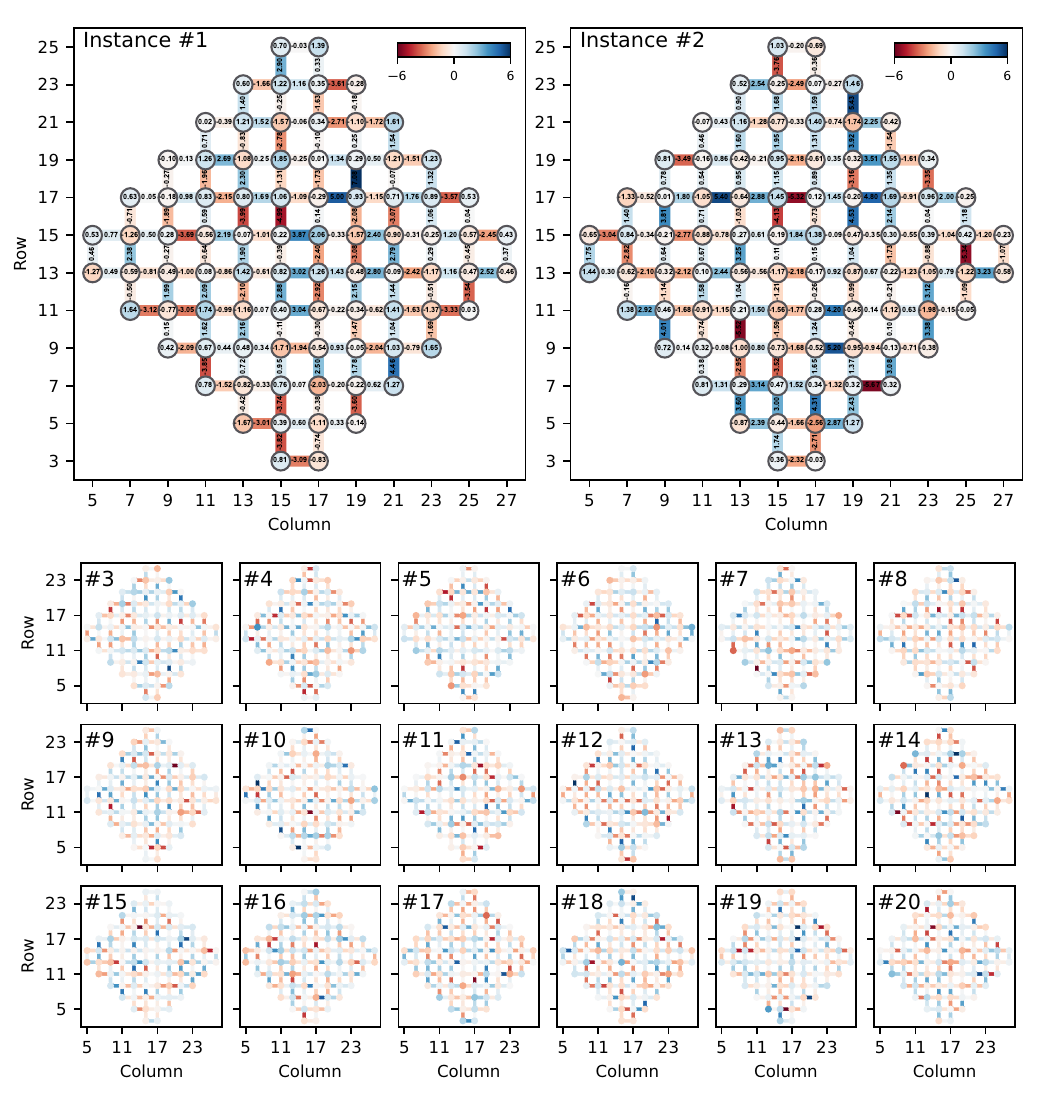}
    \caption{
    \textbf{Problem instances of $N=84$.}
    }
    \label{fig:inst_84q}
\end{figure*}

\begin{figure*}
    \centering
    \includegraphics[width=1.0\textwidth]{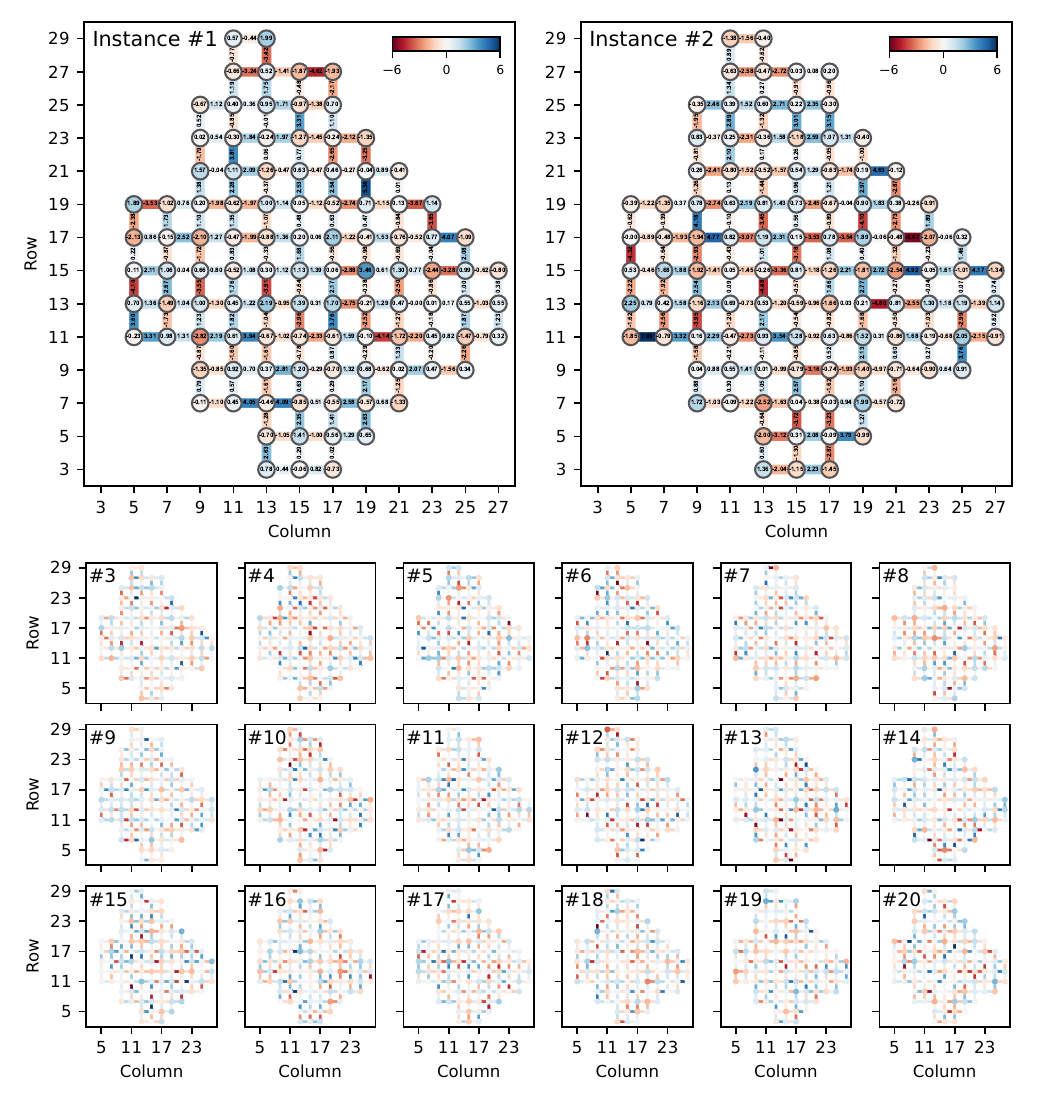}
    \caption{
    \textbf{Problem instances of $N=104$.}
    }
    \label{fig:inst_104q}
\end{figure*}

The filtering process involved generating 4000 random Ising problems according to Eq.~\ref{isingH}, solving them using a specialized SA solver (SimAn~\cite{isakov2015optimised}), and selecting the 50 instances with the highest time to solution (TTS). To further filter out instances where classical jump algorithms perform well, we selected the top 20 instances with the most challenging classical jumps as our final benchmark instances. The corresponding coupling $J_{jk}$ and local magnetic field $h_j$ of problem instances for different system sizes are presented in Fig.~\ref{fig:inst_60q},~\ref{fig:inst_84q} and \ref{fig:inst_104q}. The color of each circle indicates the magnitude of $h_j$, while the color of the connecting lines represents the magnitude of $J_{jk}$.

We excluded instances with fewer than 60 qubits for two primary reasons. First, the scaling curve is not apparent for smaller instances, only becoming stable and significant for larger instances. Second, smaller problem spaces are easier to solve without iterative methods, as the entire solution space can be explored by random initialization.

\subsection{Simulated annealing}

\begin{figure*}
\centering
\includegraphics[width=0.7\linewidth]{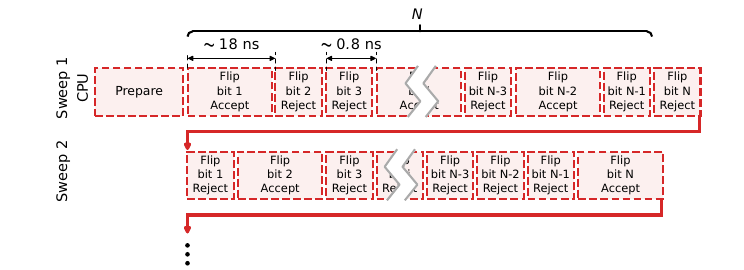}
\caption{
\textbf{Workflow of SA.} Preparation time is not included in SA runtime calculations, as this overhead does not scale with the number of runs. During each sweep, SA flips the bit one by one for all $N$ bits, and each flip and corresponding operation may take 18~ns (accept) or 0.8~ns (reject) depending on whether the flipped bitstring is accepted by the Metropolis criterion.
}
\label{fig:workflow_SA}
\end{figure*}

Simulated annealing (SA) is a powerful heuristic algorithm widely used for combinatorial optimization due to its simplicity and effectiveness. For the TTS estimation, we employed the SimAn software package~\cite{isakov2015optimised}, a highly optimized SA implementation executed on a single-core CPU (2.3 GHz) as a classical baseline. 
The effectiveness of SA is based on two facts, a high initial acceptance of any bit flips and a low escaping rate at later stages to prevent the solution from escaping the final minimum. For each problem instance, we randomly generated 10 bitstrings to calculate the average bit flip energy $\overline{|\Delta E|}$. The initial temperature $T_0$ is set to accept this  $\overline{|\Delta E|}$ with 90\% probability ($T_0 = -{\overline{|\Delta E|}}/{\ln 0.90}$), and the final temperature $T_{\text{end}}$ is set to $0.01 T_0$.
The corresponding algorithmic workflow is outlined in Algorithm~\ref{alg:sa} and Fig.~\ref{fig:workflow_SA}, and details referring to the procedure and code can be found elsewhere~\cite{isakov2015optimised}. 

\begin{algorithm}[H]
    \caption{Simulated Annealing}
    \label{alg:sa}
    \begin{algorithmic}[1]
        \State \textbf{Input:} Ising model with couplings $J_{jk}$ and fields $h_j$, initial temperature $T_0$, final temperature $T_{\text{end}}$, maximum sweeps $S_{\text{max}}$
        \State \textbf{Output:} Best solution found $\bm{s}^ {\star}$
        \State Generate initial guess solution $\bm{s}^{\circ}$
        \State $\bm{s} \gets\bm{s}^{\circ}$
        \State $\bm{s}^\star \gets\bm{s}^{\circ}$
        \State $T \gets {T_0}$
        \State Precompute and store $\Delta E_j$ list for all bits $s_j \in \bm{s}^{\circ}$
        \State For each $T$, precompute and store a list of random numbers $\{r_j\}$ corresponding to bits $\{s_j\}$, where $r_j=-T\ln u_j$ and $u_j$ is randomly chosen in $(0,1]$

        \For{sweep $=1$ to $S_{\max}$}
            \State ${T} \gets$ update according to schedule
            \State Shift the index of list $\{r_j\}$ at this temperature by a randomly generated integer
            \For{each bit $s_j$ in fixed order}
                \State Retrieve stored random number $r_j$
                \State Retrieve stored $\Delta E_j$
                \If{$\Delta E_j < r$}
                    \State Flip bit $s_j$ in $\bm{s}$ 
                    \State Update $\Delta E_k$ for $s_k \in \text{neighbors}(s_j)$
                    \If{the energy of $\bm{s}$ is lower than that of $\bm{s}^\star$}
                        \State $\bm{s}^\star \gets\bm{s}$
                    \EndIf
                \EndIf
            \EndFor
        \EndFor
        \State \Return $\bm{s}^ {\star}$ with minimal energy
    \end{algorithmic}
\end{algorithm}

Notably, this implementation of SA is highly efficient for Ising problems, incorporating several key optimizations such as:

\begin{enumerate}
\item  $\Delta E_j$ Forward computation: Precompute and store $\Delta E_j$ for each bit, with updates limited to cases where bit $s_j$ or its neighbors are flipped.

\item Precomputation and reuse of random numbers across multiple annealing runs. To improve efficiency, the Metropolis criterion is modified from $\exp(-\Delta E_j/T) < u_j
    $ (where $u_j$ is randomly chosen in $(0,1]$ and $T$ is the annealing temperature of SA) to the equivalent $\Delta E_j < r_j$ (with $r = -T\log u_j$), allowing for precomputed values of $r_j$ to be stored instead of $u_j$.
    
\item  Optimization of loops via fixed lengths. The maximum number of neighbors is specified at compile time to enable more efficient compiler unrolling.

\item  Deterministic lattice traversal. Lattice sites are updated in a predefined sequential order rather than being randomly selected.

\item  Fast random number generators: Given that random number quality is less critical for optimization algorithms (unlike high-accuracy physical simulations), fast generators (e.g., linear-congruential) are employed to reduce computational overhead.
\end{enumerate}


\subsection{Qjump time estimation}
As our current experimental setup lacks the essential elements for prompt and repeated execution of Qjump, such as rapid qubit reset, fast readout and feedback control, and in-situ data processing for local search, our analysis of the Qjump runtime can only be based on an envisioned Qjump hardware, with knowledge from the literature.
The proposed hardware, with all operations instructed by an FPGA, is able to perform a sequence of operations including qubit reset, gate operation, measurement, classical local search and comparison, and feedback control to implement gate set for the next iteration. For quibt operations, we estimate the time scales by quoting data from the literature (qubit reset time {200}~ns, gate time {40}~ns, qubit measurement time 500~ns, and feedback control time {500}~ns)~\cite{salatheLowLatencyDigitalSignal2018,acharyaSuppressingQuantumErrors2022,ai2024quantum}; for classical processing, since FPGA can be as fast as CPU, we simply use the time data recorded on CPU to estimate the time consumption of the classical local search on FPGA.

\begin{figure*}
\centering
\includegraphics[width=0.60\linewidth]{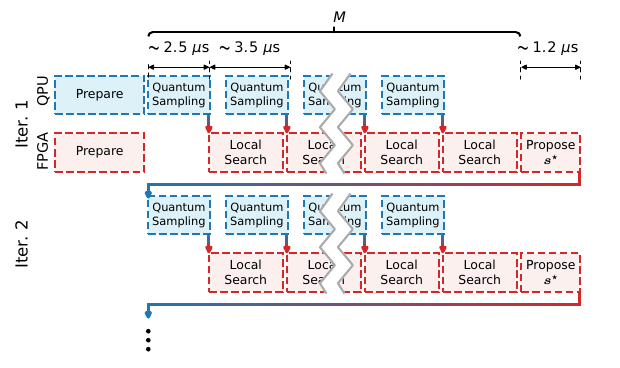}
\caption{
\textbf{Workflow of Qjump,} which is designed to run on an envisioned quantum hardware assuming that all control and data processing are on FPGA. The parallel programming ultimately minimizes the time cost of quantum operations for an optimal TTS. According to data reported in the literature, each quantum block (light blue) is speculated to take approximately 2.5~$\mu$s, which includes qubit reset (200~ns), qubit operations (33 layers and 40~ns per layer), qubit measurement (500~ns), and feedback control (500~ns)~\cite{salatheLowLatencyDigitalSignal2018,acharyaSuppressingQuantumErrors2022,ai2024quantum}. Each classical block (light red) is estimated to take approximately 3.5~$\mu$s, which consists of calculating the energy of the sampled bitstring ($\sim500$~ns) and performing roughly 18 steps of steepest descent ($\sim160$~ns per step). Preparation time is not included in Qjump runtime calculations, as this overhead does not scale with the number of iterations or runs.
}
\label{fig:workflow}
\end{figure*}

An optimized Qjump workflow is illustrated in Fig.~\ref{fig:workflow}, where the quantum sampling part (rows with blue blocks) starts one step ahead but runs almost in parallel with the classical local search part (rows with pink blocks) within $M=20$ rounds. According to this workflow, the runtime for the Qjump algorithm with $L=2$, $M=20$, and 12 iterations is estimated to be $\sim$0.78~ms , as mostly decided by the classical operations. In comparison, the runtime for the QAOA algorithm with $Q=6$ is $\sim$5.1~$\mu$s and that for the SA algorithm with 700 sweeps is estimated to be $\sim$0.30~ms, respectively.

\begin{algorithm}[H]
\caption{Greedy Local Search}
\label{alg:greedy}
\begin{algorithmic}[1]
\State \textbf{Input:}  Ising model with couplings $J_{jk}$ and fields $h_j$, system size $N$, initial guess of this iteration $\bm{s}^{\circ}$, energy of the initial guess $E^{\circ}$, candidate solution $\bm{s}$, a list $\{\Delta E^{\circ}_j\}$, inherited from the previous Qjump iteration, recording the Ising energy difference $\Delta E_j$ before and after flipping $s_j \in \bm{s}^{\circ}$
\State \textbf{Output:}  the local minima $\bm{s}^{\star}$, energy of the local minima $E^{\star}$, a list $\{\Delta E^{\star}_j\}$, recording the Ising energy difference before and after flipping $s_j \in \bm{s}^{\star}$
\State $j \gets 1$
\State $E \gets E^{\circ}$
\State $\{\Delta E_j\} \gets \{\Delta E^{\circ}_j\}$
\State $n_{\text{LS}} \gets 0$
\State /*\quad Calculate the energy of the candidate solution $\bm{s}$ \quad */
\For{$j\leq N$}
    \State $s^{\circ}_j \in  \bm{s}^{\circ}$
    \State $s_j \in  \bm{s}$
    \If{$s_j \neq s^{\circ}_j$}
        \State \sethlcolor{yellow}\hl{$E \gets E+\Delta E_j$ }
        \State  \sethlcolor{yellow}\hl{Update $\Delta E_j,~\Delta E_k \in \{\Delta E_j\}$ for $s_k \in \text{neighbors}(s_j)$ }
    \EndIf
\EndFor
\\
\State /*\quad Applying steepest descent to $\bm{s}$ \quad*/
\While{True}
    \State \sethlcolor{pink}\hl{ Find $j^{\star}$ where $\Delta E_{j^{\star}}$ is minimized in $\{\Delta E_j\}$}
    \If{$\Delta E_{j^{\star}} < 0$}
        \State Flip bit $s_j$ in $\bm{s}$
        \State \sethlcolor{yellow}\hl{ $E \gets E+\Delta E_j$}
        \State \sethlcolor{yellow}\hl{ Update $\Delta E_{j^{\star}},~\Delta E_k \in \{\Delta E_j\}$ for $s_k \in \text{neighbors}(s_{j^{\star}}) $}
        \State $n_{\text{LS}} \gets n_{\text{LS}}+1$
    \Else
        \State \textbf{break}
    \EndIf
\EndWhile
\\
\State $\bm{s}^{\star} \gets \bm{s}$
\State $E^{\star} \gets E$
\State $\{\Delta E^{\star}_j\} \gets \{\Delta E_j\}$
\State \Return $\bm{s}^{\star}$, $E^{\star}$, $\{\Delta E^{\star}_j\}$
\end{algorithmic}
\end{algorithm}

The overall Qjump runtime, as shown in Fig.~\ref{fig:workflow}, is primarily constrained by its classical computation time. This classical computation mainly involves two steps: performing a local search on each sampled result (with a time of $t_{\text{LS}}$) and proposing the optimal result from those retained in each iteration (with a time of $t_{\text{PO}}$). 
{Specifically, the local search, shown in Algorithm~\ref{alg:greedy}, decomposes into calculating the total energy of the sampled bitstring and applying steepest descent to find the local minima. These steps further break down into two fundamental operations: bit flipping and updating the energy list (with a time of $t_{\text{SF}}$), and identifying the minimum within the energy list (with a time of $t_{\text{ML}}$), which are highlighted with yellow and pink, respectively.} The runtime of local search for a candidate solution $\bm{s}$ can be formalized as:             
\begin{align*}
    t_{\text{LS}}(N,n_{\text{LS}}) = &\eta\times N \times t_{\text{SF}}
    + n_{\text{LS}}\times (t_{\text{ML}}+t_{\text{SF}}),
\end{align*}
Here, $N$ is the system size and $n_{\text{LS}}$ is the number of descent steps. $\eta$ is the average bit flip ratio for the quantum sampler. The first term represents the energy calculation time, while the second term denotes the steepest descent time.

To benchmark these individual time components ($t_\text{SF}$, $t_\text{ML}$ and $t_\text{PO}$), we executed the classical portion of Qjump on the same classical platform used for SA, with the quantum sampler replaced by a random bit flip mechanism. We performed 100 runs each on all 20 instances selected in Section~3A, each consisting of 20 iterations, and recorded the time consumption for each component. Notably, $t_{\text{PO}}$ also includes overheads for maintaining variables, copying data, and memory management within each iteration, along with an additional 200 ns delay representing the FPGA-to-QPU solution transfer time.
The average classical time costs for the system structure presented in the main text and the additional experimental data in Section~4D are summarized in Table~\ref{tab:time_cost}. 

\begin{table}[ht]
    \centering
    \begin{tabular}{l c c c}
        \toprule
        System size $N$        & 60         & 84      & 104 \\ \midrule
        $t_{\text{SF}}$ (ns) & 25.3     & 27.8      & 28.6     \\
        $t_{\text{ML}}$ (ns) & 87.7     & 107.6     & 128.2    \\ 
        $t_{\text{PO}}$ (ns) & 1378.8   & 1260.6    & 1249.2   \\ \bottomrule
    \end{tabular}
    \caption{Time consumption for individual components in the classical portion of Qjump.}
    \label{tab:time_cost}
\end{table}

For a rough approximation, we also estimated the base classical operation costs from CPU specifications.
Given the single-core CPU frequency of 2.3 GHz, one clock cycle is approximately 0.43 ns. For a system with connectivity $D\sim4$, bit flipping and updating the energy list ($t_{\text{SF}}$) involves roughly 50 clock cycles, leading to $t_{\text{SF}}\sim$20 ns. The time to identify the minimum within the energy list ($t_{\text{ML}}$) is similarly estimated to be $\sim$100 ns (at $N=104$), based on a brute-force search requiring $\sim2N$ clock cycles. We used this brute-force approach because more complex sorting algorithms offered no significant advantage for $N\sim100$ due to the overhead of maintaining additional data structures. Both values are consistent with the runtime recorded.

\begin{figure*}[ht]
    \centering
    \includegraphics[width=1.0\textwidth]{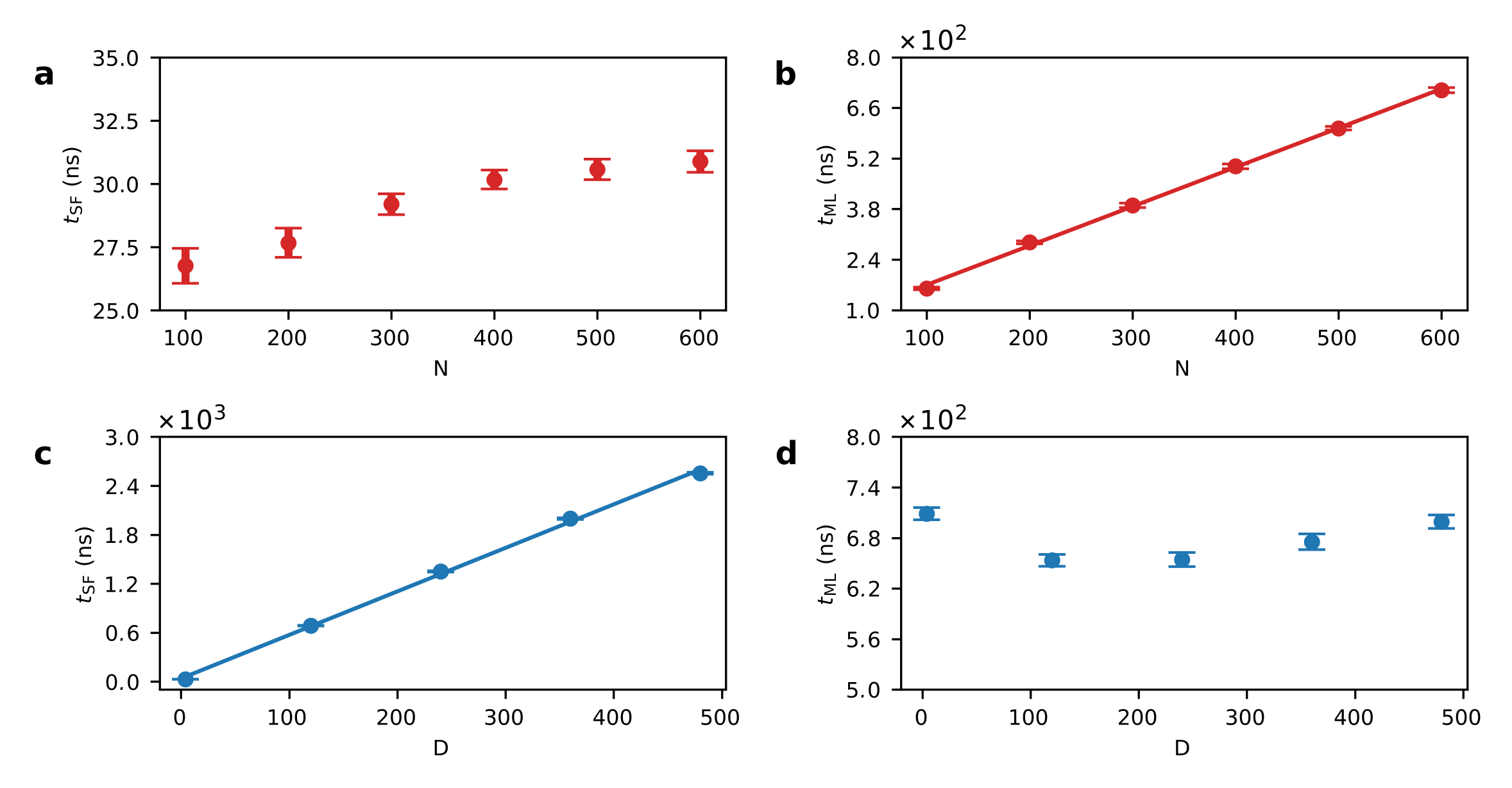}
    \caption{\textbf{Runtime benchmark and scaling behavior of individual time components in the classical portion of Qjump.} The data points represent the mean time consumption, while the error bars indicate the standard deviation over 100 runs. The solid lines show the linear fit to the data.
    }
    \label{fig:runtime}
\end{figure*}

To further analyze the time complexity of the classical portion, we extended our numerical algorithm to larger system sizes and higher degrees. We generated one random instance for each system size ($N$) ranging from 100 to 600 while keeping the degree fixed at 4. To investigate the scaling with respect to connectivity, we also generated one random instance for $N=600$ at various degrees ($D$) from $0.2N$ to $0.8N$. The time consumption for each component of the numerical algorithm on these instances is presented in Fig.~\ref{fig:runtime}.

As shown in Fig.~\ref{fig:runtime}\textbf{a} and \textbf{c}, $t_{\text{SF}}$ is primarily determined by graph connectivity $D$, rather than the system size $N$, due to an incremental gradient maintenance method yielding $ O(D)$ time complexity. For smaller system sizes, boundary effects reduce connectivity at edge qubits, leading to a smaller and more volatile $t_{\text{SF}}$. 
Conversely, $t_{\text{ML}}$ scales as $O(N)$ because it requires a traversal of all bits to find the largest improvement. Our numerical data corroborate this behavior, as shown in Fig.~\ref{fig:runtime}\textbf{b} and \textbf{d}.

\section{Experimental details}
In this section, we present the key parameters and performance metrics of our 104-qubit quantum processor (Sec.~4A). We provide a detailed description of the quantum circuits implemented in the Qjump algorithm (Sec.~4B), along with trials for different parameter settings used in Qjump (Sec.~4C). Finally, we provide additional experimental data for different qubit numbers (Sec.~4D).

\subsection{Device infomation}
\begin{figure*}
\centering
\includegraphics[width=0.95\textwidth]{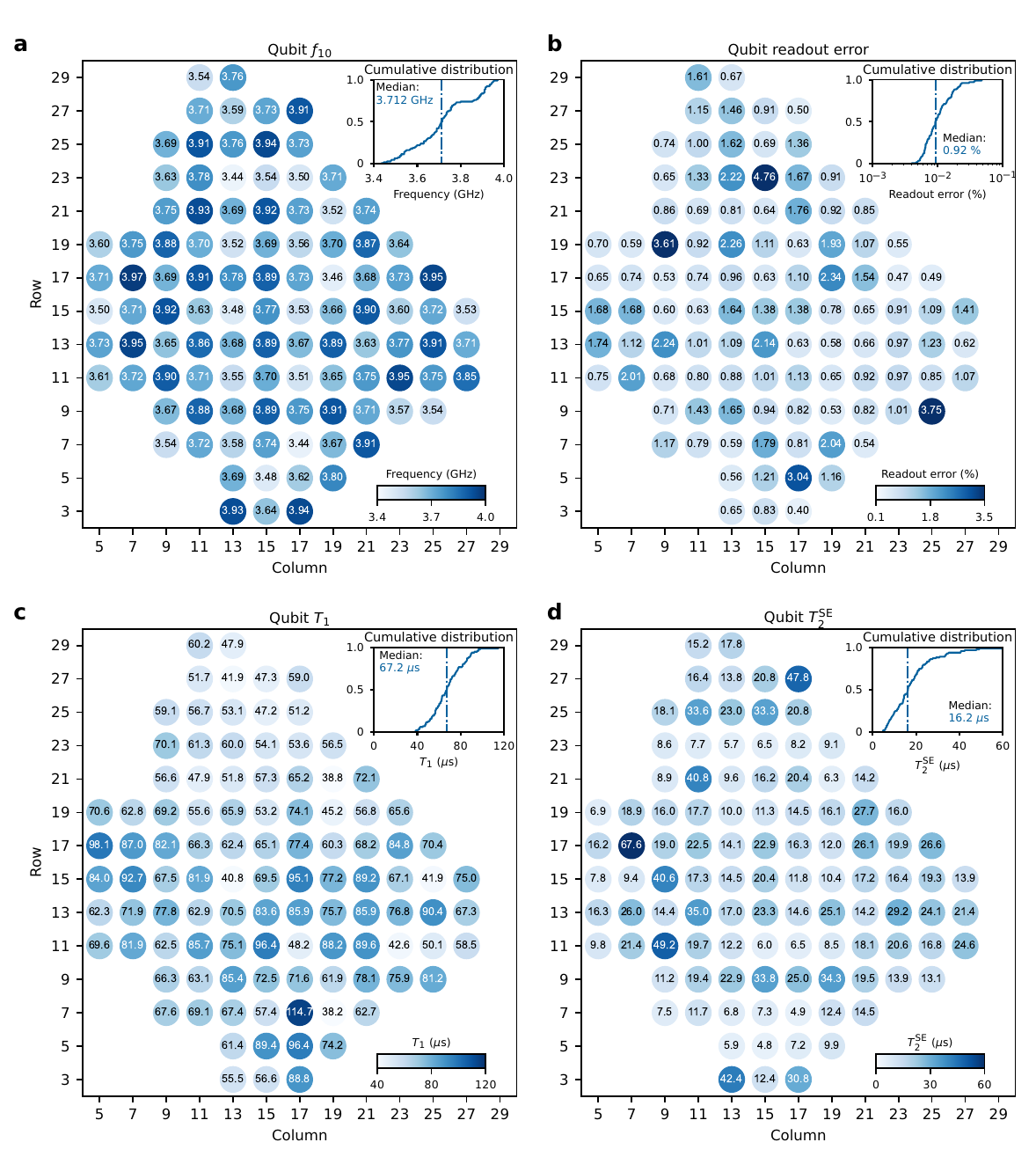}
\caption{
\textbf{Heat map of single-qubit parameters.}
\textbf{ a,}  Qubit idle frequency. 
\textbf{ b,}  Readout error averaged for qubit in $\ket{0}$ and $\ket{1}$, measured by preparing random product states on all qubits. The median for 104 qubits is 0.92\%.
\textbf{ c,}  Qubit relaxation time measured at its idle frequency. The median for 104 qubits is $67.2$~$\mu$s.
\textbf{ d,}  Qubit dephasing time measured using Hahn echo sequence at its idle frequency. The median for 104 qubits is $16.2$~$\mu$s.
}
\label{suppfig:device_info}
\end{figure*}

The wiring information and room temperature control electronics are similar to those of \cite{jin2025observation}. We summarize single-qubit parameters including idle frequency, readout error averaged for qubit in $\ket{0}$ and $\ket{1}$, energy relaxation time $T_1$, and Hahn echo dephasing time $T_2^{\text{SE}}$ in Fig.~\ref{suppfig:device_info}. The readout error was measured by initializing all qubits into random product states. 
To improve measurement fidelity, an additional microwave pulse inducing the $\ket{1} \leftrightarrow \ket{2}$ transition is applied to each qubit before the readout. 

\begin{figure*}
\centering
\includegraphics[width=0.95\textwidth]{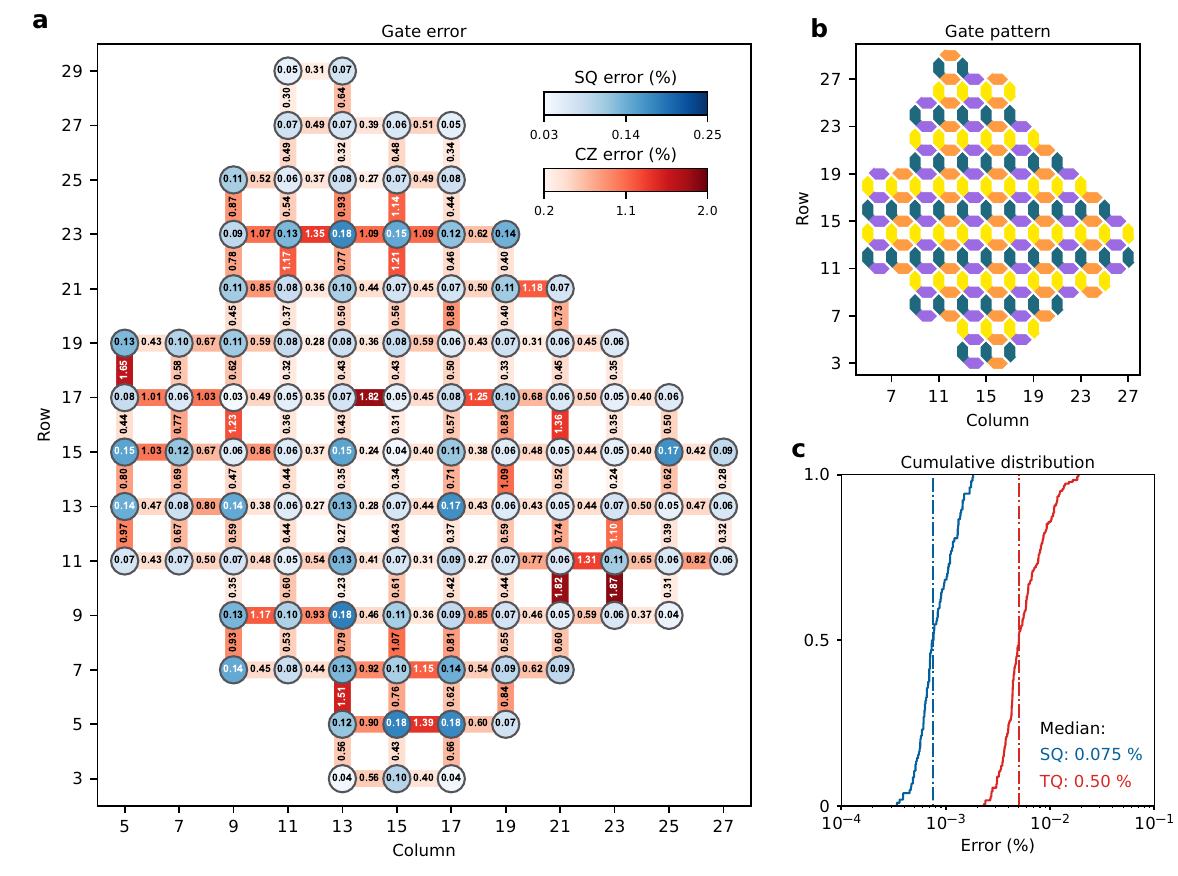}
\caption{
\textbf{Pauli errors of single- and two-qubit gates.}
\textbf{ a,}  Heatmap of Pauli errors. Gate errors are benchmarked by simultaneous XEB. Two-qubit CZ gate errors are measured based on gate patterns shown in \textbf{b}.
\textbf{ b,} Two-qubit gate patterns indexed by four different colors of the corresponding couplers. Gates with the same coupler color are benchmarked simultaneously. 
\textbf{ c,} Cumulative distribution of Pauli errors of single- and two-qubit gates. Dashed lines indicate the median values.
}
\label{suppfig:104q_gate_error}
\end{figure*}

Single-qubit gates XY rotations are realized using microwave pulses with $20$ ns duration. Two-qubit diabatic controlled-Z (CZ) gate is implemented by tuning the $\ket{11}$ and $\ket{20}$ energy levels of a qubit pair to near resonance and meanwhile activating the coupling between qubits. Figure~\ref{suppfig:104q_gate_error}b illustrates how the tunable couplers are divided into four subsets, with all corresponding CZ gates in the same subset running simultaneously. We benchmark the performance of single- and two-qubit gates by simultaneous cross-entropy benchmarking (XEB), yielding median fidelities of single- and two-qubit gates around $99.95\%$ and $99.5\%$, respectively (Fig.~\ref{suppfig:104q_gate_error}).

\subsection{Quantum circuit}

The experiment circuit, as depicted in Fig.~1\textbf{b} and Fig.~2\textbf{b} of the main text, is further compiled for hardware execution. This compiling process aims to optimize performance and mitigate errors, involving the following steps: 
\begin{enumerate}
    \item Pauli twirling: To mitigate coherent error accumulation, native CZ gates are implemented with equivalent gates that incorporate two layers of single-qubit rotations on both sides \cite{laydenQuantumenhancedMarkovChain2023,wold2025experimental}.
    \item Merging sequential single-qubit gates: Consecutive single-qubit gates are merged into a microwave XY rotation combined with a virtual Z rotation, thereby reducing the total number of single-qubit gates.
    \item Gate arrangement: Gates are arranged in the sequence to avoid the case of running single- and two-qubit gates in parallel.
    \item Dephasing suppression: Two X-gates ($\pi$ rotations around X-axis in the Bloch sphere) are applied to qubits that idle more than 150~ns.
\end{enumerate}

\subsection{Performance of the quantum sampler with different parameters}
\begin{figure}
\centering
\includegraphics[width=0.75\textwidth]{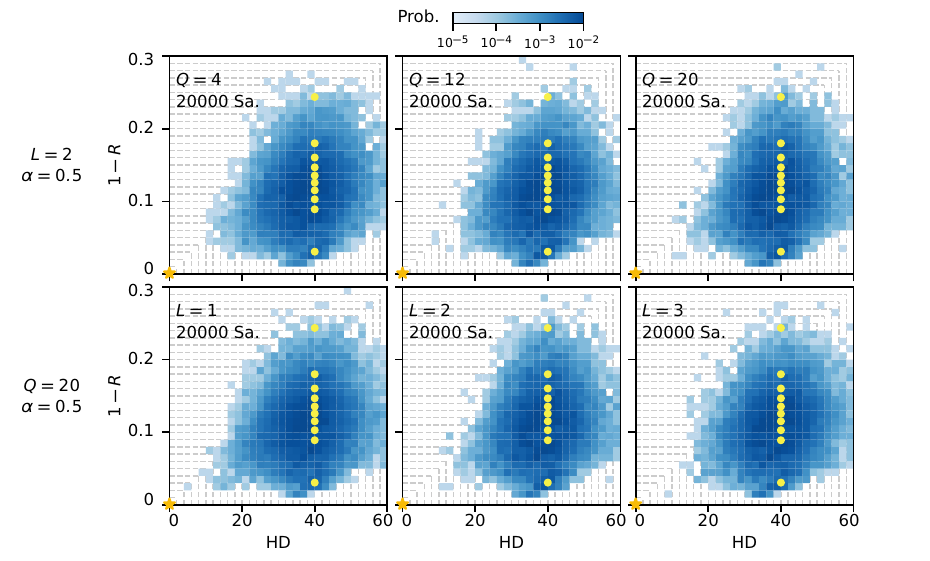}
\caption{
\textbf{Performance of the quantum sampler with different $L$ and $Q$ values.} 
Top-row panels display the sampling probability after local search ($P^{\star}$) for the quantum sampler, with varying $Q$ values at $L=2$. Bottom-row panels present $P^{\star}$ with varying $L$ values at $Q=20$. All cases start with 10 initial $\bm{s}^{\circ}$ (highlighted by yellow circles) at the Hamming distance of 40 from the global optimal bitstring (highlighted by golden star). The background gray dashed lines outline a series of square regions that approach the global minimum, where increasing proximity to the bottom-left corner indicates a closer distance to the optimum.
}
\label{suppfig:Sampler_LQ}
\end{figure}

\begin{figure}[!h]
\centering
\includegraphics[width=0.95\textwidth]{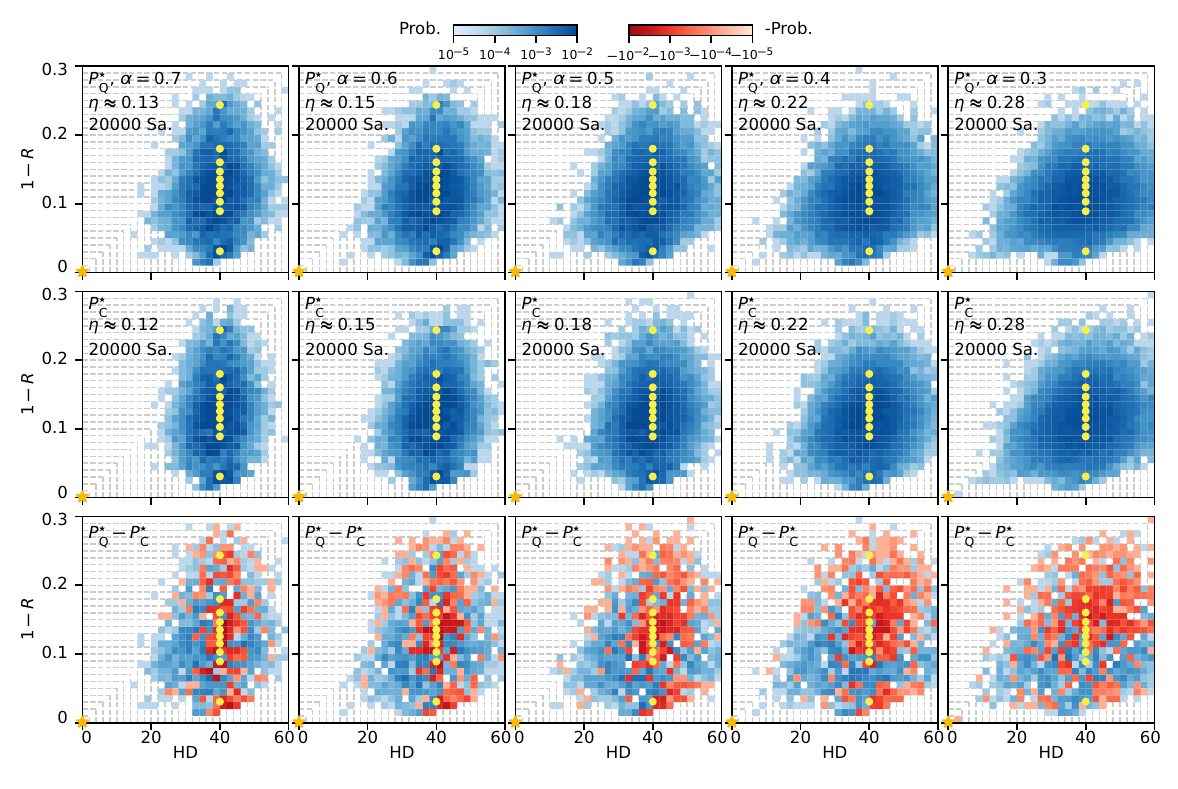}
\caption{
\textbf{Performance of the quantum sampler with different $\alpha$ values.} 
Top-row panels display the sampling probability after local search ($P^{\star}_{\text{Q}}$) for the quantum sampler, with varying $\alpha$ at $L=2$ and $Q=20$. Equivalent bit flip ratios $\eta$ are labeled. Middle-row panels present $P^{\star}_\text{C}$ by classical random sampling with comparable $\eta$ values. 
Bottom-row panels show the difference between $P^{\star}_{\text{Q}}$ and $P^{\star}_{\text{C}}$. 
}
\label{suppfig:Sampler_alpha}
\end{figure}

We experimentally tested various parameters of the quantum sampler, including $Q$ up to 20, $L$ from 1 to 3 and $\alpha \in [0.3,0.7]$ for Ising instance \#1 at $N=104$, with all configurations demonstrating similar behaviors. The sampling performance for different values of $L$ and $Q$ with a fixed $\alpha=0.5$ is presented in Fig.~\ref{suppfig:Sampler_LQ}. 
The performance of the [$2$, $20$]-sampler with $\alpha \in [0.3,0.7]$ is shown in Fig.~\ref{suppfig:Sampler_alpha}, along with comparison to classical random sampling with similar bit flip ratios. 

\subsection{Additional experimental data for different system sizes}
\begin{figure*}
\centering
\includegraphics[width=0.7\textwidth]{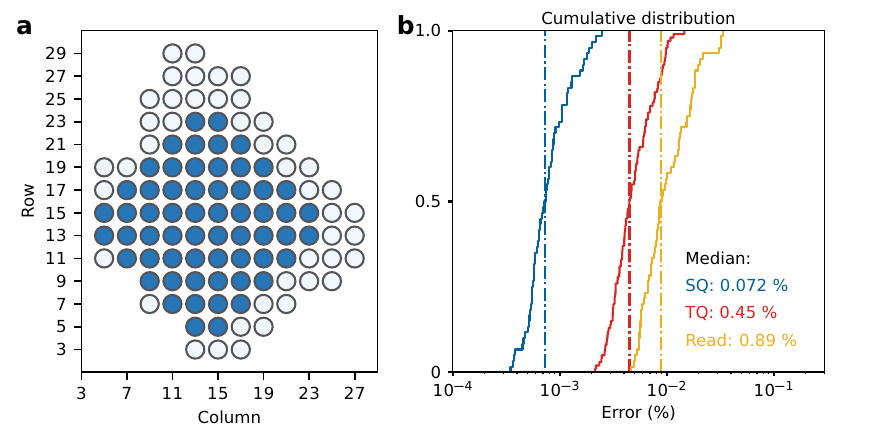}
\caption{
\textbf{Device information for $N=60$ experiment.}
\textbf{ a,} 2D lattice of qubits selected for $N=60$ experiment.
\textbf{ b,} Cumulative distribution of Pauli errors for single- and two-qubit gates, alongside the distribution of readout errors for $N=60$. Dashed lines indicate the median values.
}
\label{suppfig:qubit_info_60q}
\end{figure*}

\begin{figure*}
\centering
\includegraphics[width=0.8\textwidth]{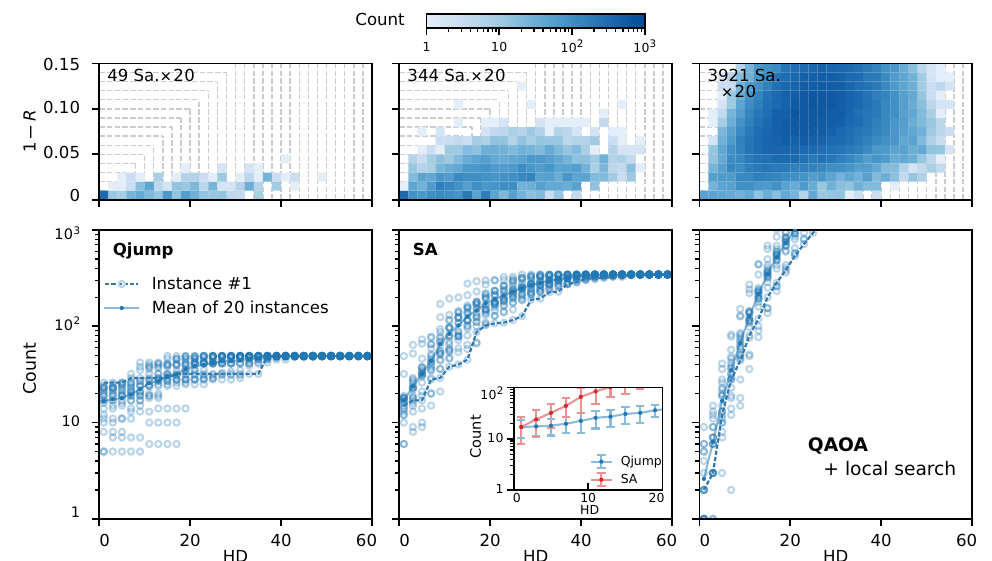}
\caption{
\textbf{Performace of Qjump, QAOA and SA for $N=60$.}
This figure summarizes the solution distributions of Qjump, SA, and QAOA across 20 problem instances.
Top row: solution count distribution over 20 problem instances for each algorithm. Bottom row: solution count for individual instances and the means as functions of HD toward the global minimum. 
For this comparison, we execute Qjump with 8 iterations at $\alpha=0.5$, SA with 200 sweeps, and QAOA with $Q=6$ and extract the lowest-energy bitstring from each run for analysis. 
Based on the computational speed of the envisioned quantum hardware and classical CPU (time per run: Qjump $\sim 0.41$~ms, SA $\sim 0.06$~ms, QAOA $\sim 5.1~\mu$s), we first determine the numbers of bitstrings that can be produced by these algorithms in repeated runs within a fixed period of 20~ms, and then run Qjump and QAOA on our superconducting processor to generate the pre-determined numbers of bitstrings for analysis, which takes a much longer time on our experimental setup. 
{For this small system size, the runtime of Qjump is primarily limited by the quantum sampling frequency rather than the classical local search.}
Mean values extracted from Qjump and SA are shown in the panel inset, with error bars representing the standard deviations over 20 instances. 
On average, Qjump samples the global minimum 0.98 times as frequently as SA.
}
\label{suppfig:alg_60q}
\end{figure*}

\begin{figure*}
\centering
\includegraphics[width=0.5\textwidth]{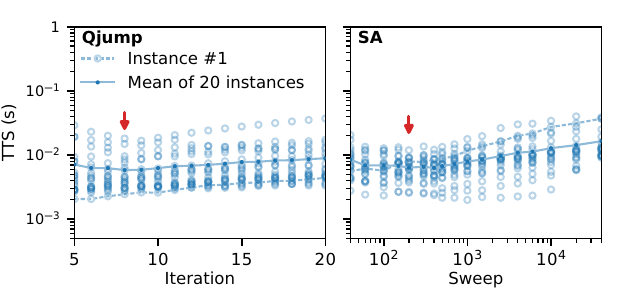}
\caption{
\textbf{TTS estimation for Qjump and SA at $N=60$.}
Each empty circle represents the TTS value for a certain problem instance obtained by running the Qjump (SA) algorithm with the specified iteration (sweep) number. Dots connected by lines indicate the mean of the data for the 20 problem instances, and the arrow points to the optimal iteration/sweep number used for the algorithmic benchmark in Fig.~\ref{suppfig:alg_60q}. TTS values for instance \#1 are highlighted by dashed lines. According to the minimal TTS values, on average, Qjump running on the envisioned quantum hardware demonstrates comparable performance to SA, with a TTS ratio of approximately 1.09.
}
\label{suppfig:TTS_60q}
\end{figure*}

\begin{figure*}
\centering
\includegraphics[width=0.7\textwidth]{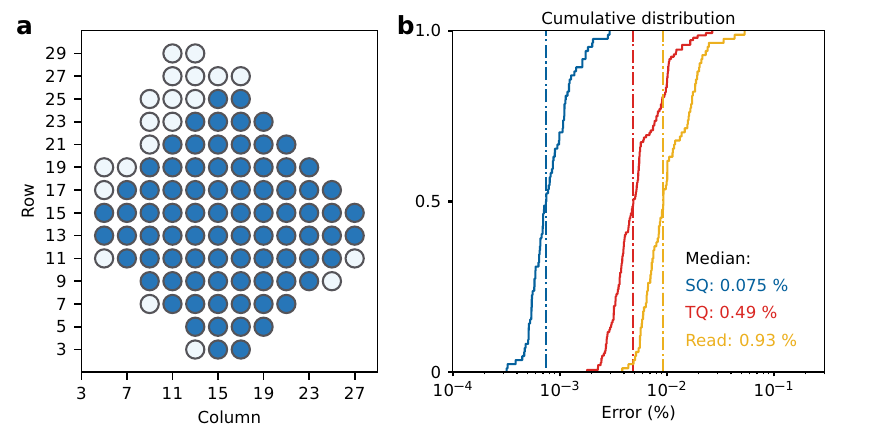}
\caption{
\textbf{Device information for $N=84$ experiment.}
\textbf{ a,} 2D lattice of qubits selected for $N=84$ experiment.
\textbf{ b,} Cumulative distribution of Pauli errors for single- and two-qubit gates, alongside the distribution of readout errors for $N=84$. Dashed lines indicate the median values.
}
\label{suppfig:qubit_info_84q}
\end{figure*}

\begin{figure*}
\centering
\includegraphics[width=0.8\textwidth]{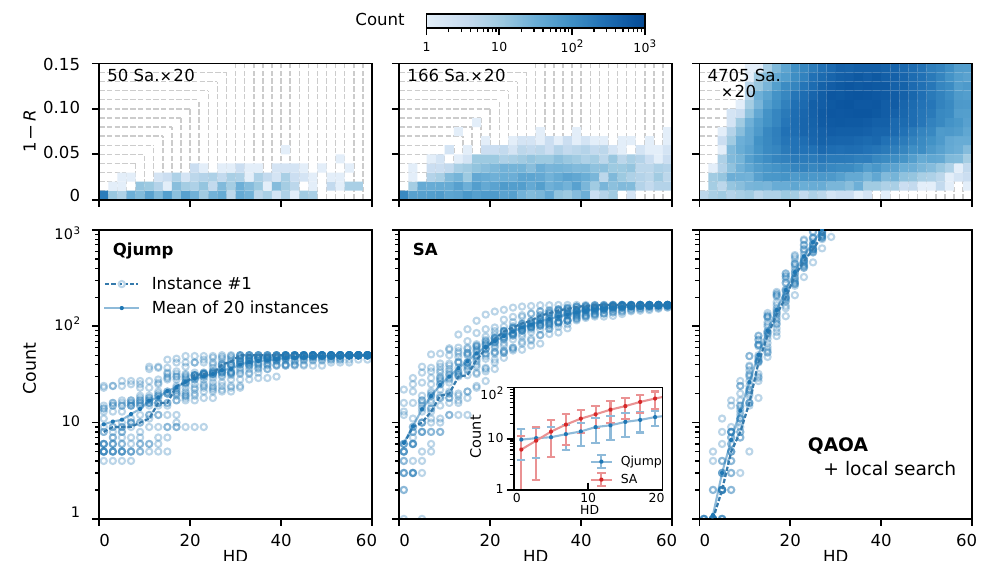}
\caption{
\textbf{Performace of Qjump, QAOA and SA for $N=84$.}
This figure summarizes the solution distributions of Qjump, SA, and QAOA across 20 problem instances.
Top row: solution count distribution over 20 problem instances for each algorithm. Bottom row: solution count for individual instances and the means as functions of HD toward the global minimum. 
For this comparison, we execute Qjump with 9 iterations at $\alpha=0.5$, SA with 400 sweeps, and QAOA with $Q=6$ and extract the lowest-energy bitstring from each run for analysis.
Based on the computational speed of the envisioned quantum hardware and classical CPU (time per run: Qjump $\sim 0.47$~ms, SA $\sim 0.14$~ms, QAOA $\sim 5.1~\mu$s), we first determine the numbers of bitstrings that can be produced by these algorithms in repeated runs within a fixed period of 24~ms, and then run Qjump and QAOA on our superconducting processor to generate the pre-determined numbers of bitstrings for analysis, which takes a much longer time on our experimental setup. 
Mean values extracted from Qjump and SA are shown in the panel inset, with error bars representing the standard deviations over 20 instances. 
On average, Qjump samples the global minimum 1.56 times more frequently than SA.
}
\label{suppfig:alg_84q}
\end{figure*}

\begin{figure*}
\centering
\includegraphics[width=0.5\textwidth]{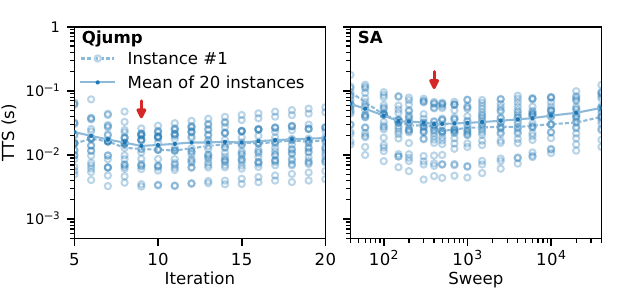}
\caption{
\textbf{TTS estimation for Qjump and SA at $N=84$.}
Each empty circle represents the TTS value for a certain problem instance obtained by running the Qjump (SA) algorithm with the specified iteration (sweep) number. Dots connected by lines indicate the mean of the data for the 20 problem instances, and the arrow points to the optimal iteration/sweep number used for the algorithmic benchmark in Fig.~\ref{suppfig:alg_84q}. TTS values for instance \#1 are highlighted by dashed lines. According to the minimal TTS values, on average, Qjump running on the envisioned quantum hardware outperforms SA by roughly a factor of 2.18.
}
\label{suppfig:TTS_84q}
\end{figure*}

In this section, we provide supplementary figures for Ising problems with $N=60$ and $N=84$. 
The qubit layouts for both systems are shown in Fig.~\ref{suppfig:qubit_info_60q} and Fig.~\ref{suppfig:qubit_info_84q}, along with the corresponding performance of single-qubit gates, CZ gates, and readout. 
Comparisons of Qjump's performance against SA and QAOA for these same system sizes are shown in Fig.~\ref{suppfig:alg_60q} and Fig.~\ref{suppfig:alg_84q}. The Qjump iteration number and the SA sweep number are optimized for each system size based on the TTS metric, and the corresponding TTS values are shown in Fig.~\ref{suppfig:TTS_60q} and Fig.~\ref{suppfig:TTS_84q}.

\clearpage
\bibliographystyle{naturemag}
\bibliography{suppRef}